\documentclass[nofootinbib,superscriptaddress,aps,prd,showkeys,noshowpacs,onecolumn,12pt]{revtex4}
\usepackage{eurosym}
\usepackage{graphics}
\usepackage{graphicx}
\usepackage{epsf}
\usepackage{bm}
\usepackage{amsmath,amssymb,amsfonts}
\usepackage{latexsym}
\usepackage{enumerate}

\def\be{\begin{equation}}
\def\ee{\end{equation}}

\makeatother

\begin{document}

\title{Testing General Relativity by Reconstructing Dark Energy and the
Growth Index with Physics-Informed Neural Networks}
\author{Andronikos Paliathanasis}
\email{anpaliat@phys.uoa.gr}
\affiliation{Institute of Systems Science, Durban University of Technology, Durban 4000,
South Africa}
\affiliation{Centre for Space Research, North-West University, Potchefstroom 2520, South
Africa}
\affiliation{Centro de Investigaci\'on, Innovaci\'on y Creaci\'on (CIIC), Universidad
Cat\'olica de Temuco, Temuco, Chile}
\affiliation{Departamento de Ciencias Matem\'{a}ticas y F\'{\i}sicas, Facultad de Ingenier%
\'{\i}a, Universidad Cat\'olica de Temuco, Temuco, Chile}
\affiliation{National Institute for Theoretical and Computational Sciences (NITheCS),
South Africa}

\begin{abstract}
We extend the Cosmo-PINN, a physics-informed neural network for
model-independent cosmological reconstruction, to reconstruct the dark
energy equation of state parameter $w_{\mathrm{DE}}(z)$ and the growth index 
$\gamma (z)$ of matter perturbations. The reconstruction is performed
through three modules: the background module, the matter growth module, and
the growth index module. The background module learns the dark energy
density $\rho _{\mathrm{DE}}(z)$ from cosmological observations, and $w_{%
\mathrm{DE}}(z)$ is reconstructed through the residual of the conservation
equation. The reconstructed background is then passed to the other two
modules, which reconstruct the observables $f(z)$ and $f\sigma _{8}(z)$
either directly from the linear perturbation equation of General Relativity
or through an independently reconstructed growth index $\gamma (z)$.
Comparing the two reconstructions on the same background provides a
consistency test of the underlying physical framework. The network is
trained on the DESI DR2 Baryon Acoustic Oscillations, Cosmic Chronometers,
Type Ia Supernovae, and redshift-space distortion data. Posterior
uncertainties for the reconstructed cosmological parameters and functions
are obtained through Hamiltonian Monte Carlo sampling. For the different
data combinations, the posterior median of the present-day dark energy
equation of state lies within the range $w_{\mathrm{DE,0}}\in \lbrack
-0.930,-0.867]$, while the phantom divide crossing takes place at $z\simeq
0.41$--$0.53$. The posterior median of the present-day growth index
reconstructed from the data lies within $\gamma _{0}\simeq 0.60$--$0.62$,
indicating suppression of the growth of cosmic structure.
\end{abstract}

\date{\today }
\maketitle

\section{Introduction}

The cosmological constant $\Lambda $, which leads to the standard $\Lambda $%
CDM model, is the simplest candidate for the description of the cosmic
acceleration \cite{rr1,Teg,Kowal,Komatsu,suzuki11}. It corresponds to a dark
energy fluid with constant energy density and pressure, and consequently a
constant equation of state parameter with the fixed value $w_{\Lambda }=-1$.
Nevertheless, the cosmological constant suffers from the fine-tuning and
coincidence problems \cite{Peri08}, while the recent Baryon Acoustic
Oscillation measurements from DESI DR2, combined with other late-time
observational measurements, indicate a preference for a dynamical character
of dark energy, in agreement with other independent analyses~\cite%
{ratra1,ratra2,ratra3}. Furthermore, the cosmological constant faces
observational tensions, the most important being the $H_{0}$ tension, where
the value obtained from the Cosmic Microwave Background differs by $\sim
5\sigma $ from that obtained from local distance measurements \cite%
{ten1,ten2,ten3,ten4}. These issues have motivated cosmologists to explore
alternative theoretical approaches for the nature of dark energy and the
origin of the cosmic acceleration, by introducing scalar fields and other
matter sources \cite{de1,de2,de3,de4,de5a,de5,de6,de7,de8,de9,de10,de11} as
dark energy candidates, or modified theories of gravity \cite%
{mod1,mod2,mod3,mod4,mod5,mod6,mod7,mod8,mod9,mode10,mod11,mod12}, where the
cosmic acceleration has a geometric origin driven by the geometric
invariants which modify the Einstein-Hilbert action.

From a phenomenological approach, various parametric forms for the dark
energy equation of state have been proposed in order to extract the
information provided by cosmological observations. The most widely used is
the Chevallier-Polarski-Linder (CPL) parametrization \cite{cpl1,cpl1a},
which can be seen as a first-order Taylor expansion of the dark energy
equation of state $w_{DE}\left( z\right) $ around the present epoch. Various
extensions and generalizations of the CPL\ model, which form the $w_{0}w_{a}$%
CDM family of models can be found in \cite%
{pp1a,pp2,pp3,pp4,pp5,pp6,pp7,pp8,pp9,pp10,pp11,pp12,pp13,pp14,pp15,pp16,pp17}
and references therein.

A natural generalization of the $w_{0}w_{a}$CDM family of models is to
expand the dark energy equation of state parameter in a suitable basis of
functions, thereby allowing an agnostic reconstruction of its evolution \cite%
{cos1,cos2,cos3,agn1}. Such an approach was considered in \cite{bas1}, where
the Pad\'{e} approximation was applied for the reconstruction of the dark
energy. The reconstruction was performed by constraining with the data the
parametric forms of $w_{DE}\left( z\right) $ for different degrees of the
expansion, and the different orders were compared through statistical
criteria. More recently, in \cite{bas2}, the authors introduced an extension
of the CPL parametrization by including higher-order terms of the Taylor
expansion. It was found that the additional degrees of freedom are penalized
by the Akaike information criterion, and that marginalizing over the
additional terms beyond the CPL parametrization weakens the evidence for
dynamical dark energy. A similar conclusion was obtained in \cite{bas3} by
using the Weighted Function Regression method.

Among the non-parametric approaches, Gaussian Processes (GP) have become one
of the most popular tools for the reconstruction of the expansion history.
In \cite{gp1}, the cosmographic parameters were reconstructed by using the
GP method, while \cite{gp2} presented a non-parametric reconstruction of the
dark energy equation of state following the same approach. The dark
energy-dark matter interaction was investigated using GP in \cite{gp3}; for
more applications of the GP in cosmology we refer the reader to \cite%
{gp4,gp5,gp6,gp7,gp15,gp16,gp8,gp9}. In GP, the reconstructed function is
characterized by a mean and a covariance function. Nevertheless, the
reconstruction is kernel dependent, and different selections for the kernel
can lead to different reconstructed histories \cite{ker1}. Moreover,
reconstructed parameters such as the dark energy equation of state follow
from higher-order derivatives of the reconstructed functions, which
increases the uncertainties. Recently, machine learning techniques have
found their way into applications in cosmology \cite{ml1,ml2,ml3}. Genetic
Algorithms \cite{ml4,ml5,ml6} and Artificial Neural Networks (ANNs) \cite%
{an1,an2,an3,an4,an5,an5a,an9}~have been widely applied for the
reconstruction of the Hubble function, the dark energy equation of state
parameter and for the reconstruction of free functions in modified theories
of gravity. Nevertheless, in the aforementioned approaches, the
reconstruction is entirely data-driven, and the reconstructed functions are
not required to satisfy the underlying physical laws governing the
cosmological dynamics.

This limitation is addressed by Physics-Informed Neural Networks (PINNs) 
\cite{pinnp}, where the physical laws are embedded in the loss function as
residual terms \cite{pinnbook}. Hence, during training, the neural network
reconstructs the physical quantities from observational data while enforcing
the underlying physical laws throughout the reconstruction domain. A PINN
for the solution of the cosmological dynamics was recently presented in \cite%
{pnn1}, while a generic gravitational consideration was introduced in \cite%
{pgr}; see also \cite{pnn3,pnn2}. A PINN framework was introduced in \cite%
{pnn6} to reconstruct the supernova absolute magnitude. Parameter estimation
for parametric dark energy models with the use of PINNs was recently
described in \cite{pnn4}. As far as the equations for the linear
cosmological perturbations are concerned, the PINN framework has been used
before in \cite{pnn5} and in \cite{pnn5a}, where in the latter study PINNs
were used for the joint reconstruction of the Hubble function and of the $%
f\sigma _{8}\left( z\right) $.

In \cite{pinnan} we introduced the Cosmo-PINN, a PINN framework for the
reconstruction of the dark energy equation of state parameter from the
background cosmological data. Within the Cosmo-PINN, the neural network
learns the dark energy density $\rho _{DE}\left( z\right) $, and the $%
w_{DE}\left( z\right) $ is reconstructed through the residual of the
conservation equation. The cosmological parameters are trained jointly with
the network, and the training is performed in two optimization stages,
followed by a Bayesian stage in which the posterior distributions of the
reconstructed functions and the cosmological parameters are calculated by
using the Hamiltonian Monte Carlo approach. The first application of the
Cosmo-PINN framework is presented in \cite{pinnbarrow}, where the varying
index of the modified Barrow entropy was reconstructed from the late-time
data.

Different theories of gravity can give rise to the same cosmic expansion
history, making it difficult to distinguish between them using background
cosmological observations alone. The growth of matter perturbations can be
used to overcome this limitation, since it provides a complementary probe to
the background expansion. Theories with the same background can predict
different matter perturbations. The growth index $\gamma \left( z\right) $
defined as $\gamma \left( z\right) =\frac{\ln f\left( z\right) }{\ln \Omega
_{m}\left( z\right) }$, has been introduced to quantify the difference among
models with the same background evolution \cite{pert1,pert2}. For the $%
\Lambda $CDM model, the growth index is approximately constant, $\gamma
_{\Lambda }\simeq \frac{6}{11}$, while values of $\gamma \left( z\right) $
close to the $\Lambda $CDM value are also obtained for parametric dark
energy models within General Relativity. However, $\gamma \left( z\right) $
has a strong dependence on the gravitational theory \cite%
{gw1,gw2,gw3,gw4,gw5,gw6}, such that it has been used to test General
Relativity on cosmological scales \cite%
{per3,aper3,per4,per5,per6,per7,per8,per9,per10}.

In this work, we extend the Cosmo-PINN to the growth of matter
perturbations, and we introduce the redshift-space distortion\ (RSD)
measurements as additional observational constraints. We introduce two
independent modules: a growth module, in which the linear perturbation
equation for the matter is introduced as a residual in the loss function and
the network derives the $f$ and $f\sigma _{8}\left( z\right) $ observables
and the growth index~$\gamma _{A}\left( z\right) $; in the second module,
the growth index $\gamma _{B}\left( z\right) $ is reconstructed
independently from the RSD data. The two modules share the same background;
hence, the comparison of the $\gamma _{A}\left( z\right) $ and $\gamma
_{B}\left( z\right) $ is a consistency test for the gravitational theory. We
consider General Relativity, where the dark matter is described by a
pressureless fluid source; thus, the new modules allow us to test General
Relativity by using the RSD data. The structure of the paper is as follows.

In Section~\ref{sec2} we present the cosmological framework, which is that
of General Relativity in an isotropic and homogeneous FLRW background, with
a dark energy fluid with a dynamical equation of state $w_{DE}(z)$, which
accounts for the late-time cosmic acceleration, and dark matter and baryons
described by pressureless fluids. We introduce the linear matter
perturbation equations and we define the growth index $\gamma (z)$. The
observational datasets used in this work are given in Section \ref{sec4}.
Furthermore, in Section~\ref{sec3} we summarize the architecture of the
PINN, together with the modifications that extend the Cosmo-PINN framework
to the reconstruction of $w_{DE}(z)$ and $\gamma (z)$. The growth index $%
\gamma \left( z\right) $ is reconstructed by two modules: from the
perturbation equations of the matter, $\gamma _{GR}\left( z\right) $, for
the underlying theory, and directly from the data of the matter
perturbations. In Section \ref{sec5} we present the reconstruction outcomes
of the training process for different combinations of the datasets.
Specifically, we investigate the effects of the supernova data on the
cosmological parameters. The reconstructed $w_{DE}(z)$ and $\gamma (z)$ are
presented with their credible intervals. Additionally, we derive the
present-day values of the cosmological parameters and of the growth index $%
\gamma _{0}$. Finally, in Section~\ref{sec6} we summarize our findings and
discuss possible extensions of the Cosmo-PINN framework.

\section{Dark Energy}

\label{sec2}

Within the framework of General Relativity, we describe the universe by a
FLRW background geometry 
\begin{equation}
ds^{2}=-N^{2}\left( t\right) dt^{2}+a^{2}\left( t\right) \left(
dx^{2}+dy^{2}+dz^{2}\right) ,  \label{cc.01}
\end{equation}%
consistent with the cosmological principle, which assumes homogeneity and
isotropy on sufficiently large scales. The function $a\left( t\right) $ is
the radius of the three-dimensional volume and \thinspace $N\left( t\right) $
is the lapse function. The comoving four-velocity is defined as $u^{\mu }=%
\frac{1}{N}\delta _{t}^{\mu }$, such that $u^{\mu }u_{\mu }=-1$, from which
we define the expansion rate by the expression~$\theta =3H,$ where $H=\frac{1%
}{N}\left( \frac{\dot{a}}{a}\right) ,~\dot{a}=\frac{da}{dt}$ is the Hubble
function. Without loss of generality, in the following we assume the gauge
for the lapse function $N\left( t\right) =1$.

Let the spacetime be filled with a cosmic fluid described by the
energy-momentum tensor 
\begin{equation}
T_{\mu \nu }=\left( \rho +p\right) u_{\mu }u_{\nu }+pg_{\mu \nu },
\label{cc.02}
\end{equation}%
where $\rho ,~$and $p$ denote the total energy density and pressure,
respectively. The equation of state parameter for the total fluid is defined
as $w_{tot}=\frac{p}{\rho }$.

The cosmic fluid consists of radiation, baryonic matter, and a dark sector
comprising dark matter and dark energy. Radiation is described by the energy
density and pressure $\rho _{r},~p_{r}=\frac{1}{3}\rho _{r}$, the baryons by~%
$\rho _{b}$,~$p_{b}=0$, and the dark sector comprises the dark matter and
the dark energy.\ Furthermore, the dark matter is considered to be given by
a pressureless fluid source, $\rho _{dm},~p_{dm}=0$, and the dark-energy
component is characterized by an energy density $\rho _{DE}$ and the
dynamical equation of state parameter $w_{DE}$, such that $p_{DE}=w_{DE}\rho
_{DE}$.

Hence, the energy-momentum tensor for the cosmic fluid is written
equivalently%
\begin{equation}
T_{\mu \nu }=\left( \rho _{m}+\rho _{b}+\frac{4}{3}\rho _{r}+\left(
1+w_{DE}\right) \rho _{DE}\right) u_{\mu }u_{\nu }+\left( \frac{1}{3}\rho
_{r}+w_{DE}\rho _{DE}\right) g_{\mu \nu }.  \label{cc.03}
\end{equation}

\subsection{Background field equations}

The Einstein field equations relate the gravitational field to the dynamics
of matter. For the FLRW background, they lead to the Friedmann equations%
\footnote{%
We consider units where $8\pi G=1$ and $c=1$.}%
\begin{align}
H^{2}& =\frac{1}{3}\left( \rho _{dm}+\rho _{b}+\rho _{r}+\rho _{DE}\right) ,
\label{cc.04} \\
\dot{H}& =-\frac{1}{2}\left( \rho _{dm}+\rho _{b}+\frac{4}{3}\rho _{r}+\left(
1+w_{DE}\right) \rho _{DE}\right) .  \label{cc.05}
\end{align}%
The Bianchi identity gives the condition\footnote{$\nabla _{\mu }$ is the
covariant derivative with respect to the Levi-Civita connection for the FLRW
background.} $\nabla _{\nu }T_{~}^{\mu \nu }=0$, which provides the
continuity equations for the cosmic fluid, that is,%
\begin{equation}
\dot{\rho}+3H\left( \rho +p\right) =0.  \label{cc.06}
\end{equation}

In the scenario where the cosmic fluid components interact only
gravitationally, from (\ref{cc.06}) it follows%
\begin{eqnarray}
\dot{\rho}_{dm}+3H\rho _{dm} &=&0,  \label{c.10} \\
\dot{\rho}_{b}+3H\rho _{b} &=&0,  \label{c.11} \\
\dot{\rho}_{r}+4H\rho _{r} &=&0,  \label{c.12} \\
\dot{\rho}_{DE}+3H\left( 1+w_{DE}\right) \rho _{DE} &=&0.  \label{c.13}
\end{eqnarray}

Integrating and replacing in (\ref{cc.04}) we find the following expression
for the Hubble function 
\begin{equation}
\left( \frac{H\left( a\right) }{H_{0}}\right) ^{2}=\Omega _{r0}\left( \frac{a%
}{a_{0}}\right) ^{-4}+\Omega _{b0}\left( \frac{a}{a_{0}}\right) ^{-3}+\Omega
_{dm0}\left( \frac{a}{a_{0}}\right) ^{-3}+\Omega _{DE0}\exp \left(
-3\int_{a}^{a_{0}}\frac{1+w_{DE}\left( \alpha \right) }{\alpha }d\alpha
\right) ,  \label{cc.07}
\end{equation}%
where $H_{0}$ is the value for the Hubble function at the present $a=a_{0}$, 
$\Omega _{i0}=\frac{\rho _{i,0}}{3H_{0}^{2}}$ denotes the energy density of
the fluid today, with the constraint%
\begin{equation}
\Omega _{dm0}+\Omega _{b0}+\Omega _{r0}+\Omega _{DE0}=1.  \label{cc.08}
\end{equation}

\subsection{Growth of matter perturbations}

In the following lines we discuss the basic elements of the evolution of the
linear growth of matter perturbations and we introduce the growth index $%
\gamma \left( z\right) $.

In our consideration, the total matter density consists of the dark matter
and the baryons, that is, $\rho _{m}=\rho _{dm}+\rho _{b}$, and because
there is no interaction term within the dark sector, the evolution of the
matter density perturbations $\delta _{m}=\frac{\delta \rho _{m}}{\rho _{m}}$%
, is governed by the second-order linear differential equation \cite%
{peebles,amend} 
\begin{equation}
\ddot{\delta}_{m}+2H\dot{\delta}_{m}-\frac{1}{2}\rho _{m}\delta _{m}=0,
\label{sc.01}
\end{equation}%
or equivalently, 
\begin{equation}
\ddot{\delta}_{m}+2H\dot{\delta}_{m}-\frac{3}{2}H^{2}\Omega _{m}\delta
_{m}=0.  \label{sc.02}
\end{equation}%
where $\Omega _{m}=\frac{\rho _{m}}{3H^{2}}$ denotes the energy density of
the total matter.

The growth rate of clustering is defined as $f\left( a\right) =\frac{d\ln
\delta _{m}}{d\ln a}$. In terms of the new variables, equation (\ref{sc.01})
reads%
\begin{equation}
\frac{df}{d\ln a}+f^{2}+\frac{1}{2}\left( 1-3w_{eff}\left( a\right) \right)
f-\frac{3}{2}\Omega _{m}(a)=0,  \label{sc.03}
\end{equation}%
where $w_{eff}\left( a\right) $ is the equation of state parameter for the
cosmic fluid, that is, 
\begin{equation}
w_{eff}=\frac{1}{3}\Omega _{r}+w_{DE}\Omega _{DE}.
\end{equation}%
In the late-time epoch of the universe, radiation can be neglected; thus,
the latter equation reads%
\begin{equation}
\frac{df}{d\ln a}+f^{2}+\frac{1}{2}\left( 1-3w_{DE}\left( a\right) \left(
1-\Omega _{m}\left( a\right) \right) \right) f-\frac{3}{2}\Omega _{m}(a)=0
\label{sc.04}
\end{equation}%
from which we observe that $w_{DE}$ affects the clustering of the matter
component through the background evolution. The first-order ordinary
differential equation (\ref{sc.04}) is of Riccati type.

Consider now the growth index $\gamma \left( a\right) $, defined as 
\begin{equation}
f=\Omega _{m}\left( a\right) ^{\gamma \left( a\right) },  \label{sc.05a}
\end{equation}%
then $\gamma \left( a\right) $ satisfies the equation%
\begin{equation}
\frac{d\gamma }{d\ln a}\ln \Omega _{m}+\Omega _{m}^{\gamma }+3w_{DE}\left(
a\right) \left( 1-\Omega _{m}\right) \left( \gamma -\frac{1}{2}\right) +%
\frac{1}{2}-\frac{3}{2}\Omega _{m}^{1-\gamma }=0.  \label{sc.05}
\end{equation}

In this set of variables, for a slowly varying equation of state parameter $%
\frac{dw_{DE}}{d\Omega _{m}}<<\frac{1}{1-\Omega _{m}}$, i.e. $w_{DE}\simeq
w_{DE,0}$, it follows to first approximation \cite{pert1,pert2}~ 
\begin{equation}
\gamma \simeq \frac{3\left( 1-w_{DE,0}\right) }{5-6w_{DE,0}}\text{.}
\label{sc.06}
\end{equation}%
where for $\Lambda $CDM, in which $w_{DE,0}=-1$, we calculate $\gamma \simeq 
\frac{6}{11}$, such that $f\left( a\right) =\Omega _{m}^{\frac{6}{11}} $.

The growth index $\gamma \left( a\right) $ can be used as a phenomenological
parameter to characterize deviations from the $\Lambda $CDM, as well as from
the case of a constant equation of state parameter. In the literature,
various phenomenological parametrizations have been proposed to describe the
evolution of the growth index. Some models that have been considered beyond
the constant $\gamma \left( z\right) =\gamma _{0}$ are, for instance, $%
\gamma \left( z\right) =\gamma _{0}+\gamma _{1}\frac{z}{1+z}$,$~\gamma
\left( z\right) =\gamma _{0}+\gamma _{1}ze^{-z}$, $\gamma \left( z\right)
=\gamma _{0}+\gamma _{1}\left( 1-\Omega _{m}\left( z\right) \right) ~$and
others \cite{per3,aper3,per4,per5,per6,per7,per8,per9,per10}.

\section{Cosmological Data}

\label{sec4}

In our analysis, we combine background cosmological observations with the
Redshift-Space Distortion (RSD) measurements from galaxy redshift surveys.
From these data sets we can constrain both the background expansion history
and the growth of matter perturbations.

The RSD data probe the growth of large-scale structure and constrain the
evolution of matter perturbations. We make use of the catalogues presented
in \cite{rsd1,rsd2}, which provide measurements of the growth rate $f\left(
z\right) $ and of $f\sigma _{8}\left( z\right) $ at given redshifts, where $%
\sigma _{8}\left( z\right) $ is the root-mean-square amplitude of matter
fluctuations in spheres with radius $8h^{-1}Mpc$, that is, $\sigma
_{8}\left( z\right) =\sigma _{8,0}\frac{\delta _{m}\left( z\right) }{\delta
_{m}\left( 0\right) }$. The $f\left( z\right) $ measurements cover the
redshift range $0.013\leq z\leq 1.40$, while the $f\sigma _{8}\left(
z\right) $ measurements span $0.02\leq z\leq 1.944$.

For the background constraints, we use Cosmic Chronometers (CC), Type Ia
Supernovae (SNIa), and Baryon Acoustic Oscillations (BAO).

The CC catalogue provides measurements of the Hubble parameter at redshifts\
in the range $0.09\leq z\leq 1.965$, from old-aged galaxies \cite{co01,co02}%
. The likelihood is defined with the full covariance matrix\footnote{%
https://gitlab.com/mmoresco/CCcovariance} as introduced in \cite{more11}.

Moreover, we consider the BAO measurements as provided by the Dark Energy
Spectroscopic Instrument Data Release 2 (DESI DR2) \cite{desi1,desi2}. The
catalogue includes seven bins of measurements for the comoving angular
distance, the volume-averaged distance and the Hubble distance normalized by
the sound horizon at the baryon drag epoch $r_{drag}$, that is, $%
D_{M}r_{drag}^{-1}$, $D_{V}r_{drag}^{-1}$ and $D_{H}r_{drag}^{-1}\,$.

Finally, we consider the PantheonPlus (PP) \cite{pp1}, Union3.0 (U3) \cite%
{union3} and DES-Dovekie (DD) \cite{desd} SNIa catalogues. The PP sample
contains 1550 SNIa observations in the redshift range $10^{-3}<z<2.27$,
while the U3 includes 2087 SNIa covering a comparable redshift range. DESD
is the most recent catalogue, which follows from the reanalysis of the
five-year Dark Energy Survey supernova program (DES-SN5YR), and includes
1820 SNIa events in the redshift range$~z<1.13$. We consider the PP sample
with and without the Cepheid calibration from SH0ES.

\section{Physics-Informed Neural Network}

\label{sec3}

The architecture of the PINN adopted in this study is based on the
Cosmo-PINN framework introduced in \cite{pinnan}. The Cosmo-PINN is designed
to reconstruct the dark energy equation of state parameter $w_{DE}(z)$ from
background cosmological observations, with the underlying physical laws
imposed as constraints on the neural network rather than being solved
independently by conventional numerical methods. This neural network
architecture ensures that the reconstructed cosmological history remains
consistent with the gravitational theory at every point of the redshift
domain.

Within the Cosmo-PINN, the training provides a single optimal solution that
minimizes the loss function, the maximum a posteriori (MAP) solution.
Uncertainties of the reconstructed functions are derived in a Bayesian
phase. For the derivation of the posterior, we perform sampling with the
Hamiltonian Monte Carlo method, starting from the MAP solution. In this
work, we extend the architecture of the Cosmo-PINN by introducing a
dedicated growth module, such that the network provides the evolution of the
matter density perturbations $\delta _{m}$, which is constrained by the
linear growth equation, and the growth rate $f(z)$ and the combination $%
f\sigma _{8}(z)$, so that the RSD measurements can be used as data for the
reconstruction process. Moreover, the extended network introduces an
independent reconstruction of the growth index $\gamma \left( z\right) $ and
of the growth observables through $f=\Omega _{m}^{\gamma }$ from the RSD
data, without solving the master equation (\ref{sc.01}) for the
perturbations; we shall refer to this as the growth index module.

The architecture of the PINN is schematically illustrated in Figure \ref%
{fig0}. The Cosmo-PINN consists of three stages: an Adam optimization stage,
an L-BFGS optimization stage, and a Bayesian inference stage, which provides
the posterior space for the reconstructed cosmological functions and
cosmological parameters. 
\begin{figure}[t]
\centering\includegraphics[width=1\textwidth]{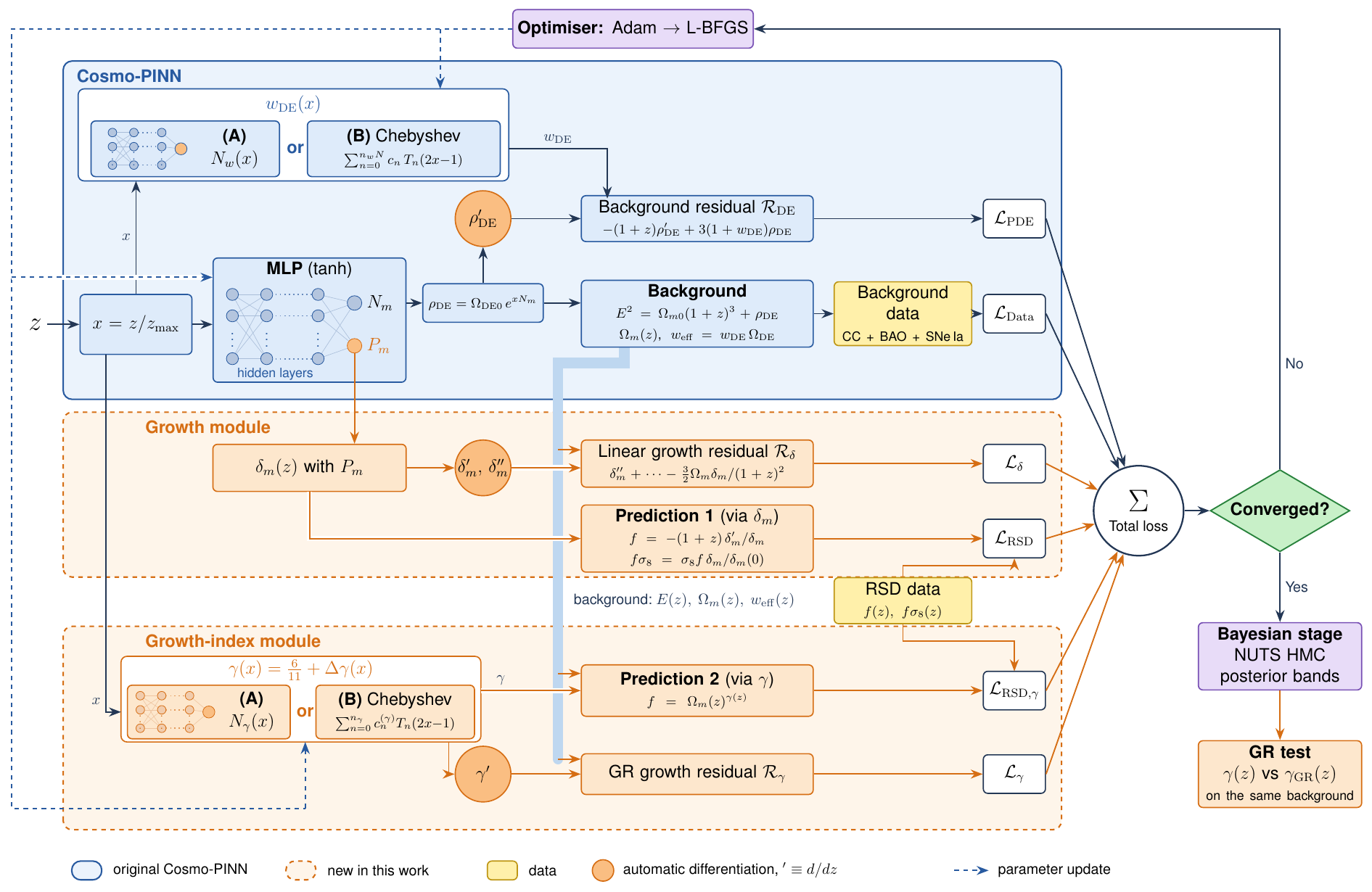}
\caption{Architecture of the extended Cosmo-PINN. The original framework for
the reconstruction of $w_{\mathrm{DE}}(z)$ from background data is shown in
blue. The growth module for $\protect\delta _{m}(z)$ and the growth-index
module for $\protect\gamma (z)$ introduced in this work are shown in orange.}
\label{fig0}
\end{figure}

\subsection{Cosmo-PINN Background}

Within the Cosmo-PINN framework, the redshift input is mapped to the unit
interval via the transformation $x=\frac{z}{z_{\max }}$, in which $x\in %
\left[ 0,1\right] $, and $z_{\max }$ is the largest value for the redshifts
of the data sets, that is, $z_{\max }=2.33$. The multilayer perceptron (MLP)
provides the function $N_{DE}(x)$, which is related to the normalized dark
energy density \cite{pinnan} 
\begin{equation}
\bar{\rho}_{DE}\left( z\right) =\Omega _{DE0}\,e^{xN_{m}(x)}\,,
\end{equation}%
where the Hubble rate is given by the expression\footnote{%
Radiation can be neglected in the background evolution at late times.}%
\begin{equation}
E^{2}(z)\equiv \left( \frac{H(z)}{H_{0}}\right) ^{2}=\Omega _{m0}(1+z)^{3}+%
\bar{\rho}_{DE}(z).
\end{equation}%
The definition of the $\bar{\rho}_{DE}\left( z\right) $ with the exponential
function is necessary in order to impose $\rho _{DE}\left( z\right) \geq $ $%
0 $ during the training.

The dark energy equation of state parameter is trained via the residual \cite%
{pinnan}%
\begin{equation}
\mathcal{R}_{PDE}:(1+z)\,\frac{d\bar{\rho}_{DE}}{dz}-3\left(
1+w_{DE}(x)\right) \,\bar{\rho}_{DE}(x).
\end{equation}%
For the representation of the $w_{DE}\left( x\right) $ within the PINN, we
employ two different approaches. We expand $w_{DE}\left( x\right) $ in a
Chebyshev polynomial basis \cite{cheb1} 
\begin{equation}
w_{DE}(x)=\sum_{n=0}^{N}c_{n}^{(w)}\,T_{n}(2x-1),
\end{equation}%
in which the coefficients are initialized at the $\Lambda $CDM point, $%
c_{0}^{(w)}=-1$ and $c_{n}^{(w)}=0$ for $n\geq 1$, so training begins from $%
w_{DE}=-1$. In the second representation form, we allow $w_{DE}\left(
x\right) $ to be described by an independent neural network, initialized at
the cosmological constant limit. As discussed in \cite{pinnan}, the two
representations lead to statistically equivalent reconstructions of $%
w_{DE}\left( z\right) $ for a sufficiently large number $N$ of the Chebyshev
expansion. The Chebyshev representation should not be confused with a
parametric form of $w_{DE}\left( z\right) $, but rather a series
approximation of the function $w_{DE}\left( x\right) $ that follows from the
trained dark energy density $\bar{\rho}_{DE}$ via the residual~$\mathcal{R}%
_{PDE}$.

The SNIa compilations, the CC and the BAO catalogues are used to constrain
the background, and the cosmological parameters $H_{0},~\Omega
_{m0},~r_{drag} $ are provided jointly with the Chebyshev coefficients $%
c_{n}^{(w)}$ for the dark energy equation of state parameter. The
coefficients are initialized at the $\Lambda $CDM point, $c_{0}^{(w)}=-1$
and $c_{n}^{(w)}=0$ for $n\geq 1$, so training begins from $w_{DE}=-1$.
Hence, the network is trained to explore deviations from the $\Lambda $CDM
limit.

Additionally, the network provides a second output, the $P_{m}\left(
x\right) $ which is used in the training of the growth sector. The physical
quantities $\Omega _{m}\left( z\right) =\frac{\Omega _{m0}\,(1+z)^{3}}{%
E^{2}(z)}$ and $w_{eff}=w_{DE}\left( z\right) \left( 1-\Omega _{m}\left(
z\right) \right) $ are passed to the growth sector as background quantities.
Thus, all growth quantities are derived on the same background.

\subsection{Growth module $\protect\delta _{m}$}

\label{sec3.2}

For the growth module, the output $P_{m}\left( x\right) $ from the main
neural network is used to define $\delta _{m}\left( z\right) $ through the
formula%
\begin{equation}
\delta _{m}\left( x\right) =\frac{1+z_{\max }}{1+xz_{\max }}\left( 1+\kappa
\left( 1-x\right) \right) \left( 1+\left( 1-x\right) ^{2}P_{m}\left(
z\right) \right) ,
\end{equation}%
where now\footnote{%
We remark that the value $\delta _{m}\left( z_{\max }\right) $ is not
essential, since the observables are $f(z)$ and $f\sigma _{8}(z)$, and they
are invariant under scaling transformations of the $\delta _{m}\left(
z\right) $.} $\delta _{m}\left( z_{\max }\right) =1,~$and for $z$ close to
the matter epoch $\delta _{m}\left( z\right) \simeq a.$ The parameter $%
\kappa $ is related to the value of the growth rate at the matter epoch as $%
f\left( z_{\max }\right) =1+\kappa \left( 1+\frac{1}{z_{\max }}\right) $. $%
f\left( z_{\max }\right) $ is a free parameter of the model, and it is
trained by the RSD data; $f\left( z_{\max }\right) =f_{\ast }$ is the
initial condition of the differential equation (\ref{sc.03}).

In order for the network to satisfy the physical law, the following residual
term is introduced within the loss function%
\begin{equation}
\mathcal{R}_{PERT}(z):\left( 1+z\right) ^{2}\delta _{m}^{\prime \prime }+%
\frac{1}{2}\left( 1+z\right) \left( 1+3w_{eff}\left( z\right) \right) \delta
_{m}^{\prime }-\frac{3}{2}\Omega _{m}\left( z\right) \delta _{m}
\end{equation}%
with a varying weight $\lambda _{\delta }$.

The theoretical observables $f(z)$ and the $f\sigma _{8}(z)$ are derived and
constrained using the RSD data via the loss term 
\begin{equation}
\mathcal{L}_{\mathrm{RSD}}=\frac{\lambda _{\mathrm{RSD}}}{N_{\mathrm{RSD}}}%
\chi _{\mathrm{RSD}}^{2}~;~\chi _{\mathrm{RSD}}^{2}=\chi _{\mathrm{f}%
}^{2}+\chi _{\mathrm{f\sigma }_{8}}^{2}
\end{equation}%
Furthermore, following the construction of the Cosmo-PINN, weak penalties on
the first and second derivatives of the trained function $P_{m}\left(
z\right) $ are included in the loss function.

Furthermore, from the reconstructed $\delta _{m}$, we calculate the growth
index $\gamma _{GR}\left( z\right) $.

\subsection{Growth index module $\protect\gamma \left( z\right) $}

\label{sec3.3}

For the reconstruction of the growth index, we follow the same approach as
for the reconstruction of the dark energy equation of state parameter, by
using a Chebyshev expansion, or a network output.

We employ a Chebyshev expansion with the basis%
\begin{equation}
\gamma (x)=\frac{6}{11}+\sum_{n=0}^{N_{\gamma }}c_{n}^{\left( \gamma \right)
}T_{n}(2x-1).
\end{equation}%
where $c_{0}^{\left( \gamma \right) }$ is initialized around the $\gamma
_{\Lambda }=\frac{6}{11}$ value. With this choice, we are able to track
deviations from the $\Lambda $CDM. Moreover, it is possible to express $%
\gamma \left( x\right) $ from a third neural network, similarly to the
representation of $w_{DE}\left( x\right) $, as presented in Fig. \ref{fig0}.

Furthermore, we independently derive the growth rate $f_{\gamma }(z)=\Omega
_{m}(z)^{\gamma (z)}$ and the corresponding $f\sigma _{8,\gamma }\left(
z\right) $ observable directly from the reconstructed $\gamma \left(
z\right) $, where $\Omega _{m}\left( z\right) $ is given by the background,
and now 
\begin{equation*}
\sigma _{8,\gamma }\left( z\right) =\sigma _{8,0}\exp \!\left[ -\int_{0}^{z}%
\frac{f_{\gamma }(z^{\prime })}{1+z^{\prime }}\,dz^{\prime }\right] .
\end{equation*}%
Finally, the loss term is introduced%
\begin{equation}
\mathcal{L}_{\mathrm{RSD},\gamma }=\lambda _{\mathrm{RSD,\gamma }}\left( 
\frac{\chi _{\mathrm{f}_{\gamma }}^{2}}{N_{\mathrm{f}_{\gamma }}}+\frac{\chi
_{\mathrm{f\sigma }_{8,\gamma }}^{2}}{N_{\mathrm{f\sigma }_{8,\gamma }}}%
\right) ,
\end{equation}%
in order to drive the reconstruction from the data.

The RSD data constrain the function $\gamma \left( z\right) $; nevertheless,
the growth index within General Relativity should satisfy the differential
equation (\ref{sc.05}).\ We impose the physical law as a residual term%
\begin{equation}
\mathcal{R}_{\gamma }(z):-\left( 1+z\right) \frac{d\gamma }{dz}\ln \Omega
_{m}+\Omega _{m}^{\gamma }+3w_{DE}\left( a\right) \left( 1-\Omega
_{m}\right) \left( \gamma -\frac{1}{2}\right) +\frac{1}{2}-\frac{3}{2}\Omega
_{m}^{1-\gamma },
\end{equation}%
within the loss function with a constant weight $\lambda _{\gamma }$,
together with a soft penalty on the first derivative of the $\gamma \left(
z\right) \,$.

This module introduces two loss components: the $\mathcal{L}_{\mathrm{RSD}%
,\gamma }$ from the data, and the $\mathcal{L}_{\gamma }=\lambda _{\gamma
}\left( \mathcal{R}_{\gamma }(z)\right) ^{2}$ of the residual. In the first
two stages, the network is optimized to provide the function $\gamma \left(
z\right) $ which minimizes the sum $\mathcal{L}_{\mathrm{RSD},\gamma }+%
\mathcal{L}_{\gamma }$, and not necessarily the individual loss components,
since the data do not necessarily satisfy the given physical law. If the RSD
data do not support the given physical law, there is no function $\gamma
\left( z\right) $ that can minimize both loss components. Nevertheless, in
the Bayesian stage, only the $\mathcal{L}_{\mathrm{RSD},\gamma }$ is used to
train the growth index $\gamma \left( z\right) $. For the reconstructed
background, we independently calculate the function $\gamma _{\mathrm{GR}}(z)
$ from the differential equation (\ref{sc.05}). The deviation between the
two independent reconstructed functions $\gamma \left( z\right) $ and $%
\gamma _{\mathrm{GR}}\left( z\right) $ can be seen as a test of whether the
physical law (\ref{sc.05}) is supported by the data, that is, whether our
cosmological framework of General Relativity with a pressureless description
for the dark matter is in agreement with the given observations.

\subsection{Training and Posterior Uncertainties}

The network consists of three hidden layers with 64 neurons each, and as the
activation function we consider $tanh$. The training proceeds in two
sequential phases: the Adam optimizer and the L-BFGS optimizer epochs. The
Adam optimizer is employed for 15000 epochs and the L-BFGS optimizer is
applied for up to 200 epochs, with early stopping once the total loss term
does not decrease. The individual loss components differ by orders of
magnitude in the size of their gradients \cite{gradnorm}. Hence, in order to
prevent a specific residual from driving the reconstruction, we employ the
GradNorm method during the first 5000 epochs of the Adam optimizer to adjust
the loss weights of the data sets and the residuals. After this epoch, the
weights are frozen and remain fixed for the rest of the training.

Furthermore, when the solution has converged, the Bayesian stage takes
place. The Hamiltonian Monte Carlo (HMC) \cite{hmc1} approach with the
No-U-Turn Sampler (NUTS) \cite{hmc2} is considered, which allows us to
sample the posterior distribution over the free cosmological parameters,
namely $H_{0},~\Omega _{m0},~r_{drag},~\sigma _{8,0}$ and $f_{\ast }=f\left(
z_{\max }\right) $, the Chebyshev coefficients $c_{N}^{w}$ of $w_{DE}\left(
z\right) $, the coefficients $c_{N}^{\gamma }$ for the $\gamma \left(
z\right) $ and the weights of the final network layer around the MAP
solution, which provides the outputs $N_{m}\left( x\right) $ and $%
P_{m}\left( x\right) $. For each posterior sample, the functions $\rho
_{DE}(z)$, $E(z)$, $\Omega _{m}(z),$ are derived, while $\gamma _{GR}\left(
z\right) $ is calculated independently from equation (\ref{sc.05}). For the
reconstructed functions $w_{DE}\left( z\right) $ and $\gamma \left( z\right) 
$ we consider the Chebyshev expansions of order $N_{w}=5$ and $N_{\gamma }=3$
respectively. In the analysis presented in \cite{pinnan} it was found that
there is no significant improvement in the reconstruction when we consider
more terms in the expansion for the $w_{DE}\left( z\right) $, in agreement
with the result obtained in \cite{mpp1} for a polynomial basis for the
reconstruction of the dark energy equation of state parameter. Moreover, for
the $\gamma \left( z\right) $ we assume a smaller number, since the total
number of RSD data points is relatively small.

The Gaussian priors for the free cosmological parameters and the Chebyshev
coefficients are presented in Tables \ref{priors0} and \ref{priors2}
respectively. For the Chebyshev coefficients \cite{cheb2}, we consider
scalable priors, in order for the lower-order coefficients to have strong
constraints and to avoid degeneracies as well as oscillatory behaviour; for
more details, we refer to the discussion in \cite{lasso}.

\begin{table}[tbp] \centering%
\caption{Gaussian priors for the cosmological parameters.}%
\begin{tabular}{cc}
\hline\hline
\textbf{Parameter} & \textbf{Prior} \\ \hline
$\mathbf{H}_{0}\mathbf{\ }$ & $\mathcal{N}(68.0,\,2.0^{2})$ \\ 
$\mathbf{\Omega }_{m0}$ & $\mathcal{N}(0.30,\,0.03^{2})$ \\ 
$\mathbf{r}_{drag}$ & $\mathcal{N}(147,\,3.0^{2})$ \\ 
$\mathbf{\sigma }_{8,0}$ & $\mathcal{N}(0.81,\,0.10^{2})$ \\ 
$\mathbf{f}_{\star }$ & $\mathcal{N}(1.0,\,0.2^{2})$ \\ \hline\hline
\end{tabular}%
\label{priors0}%
\end{table}%

\begin{table}[tbp] \centering%
\caption{Gaussian priors for the Chebyshev coefficients. The priors are
centered on the $\Lambda$CDM and on the constant growth index $\gamma=6/11$.}%
\begin{tabular}{cccc}
\hline\hline
\textbf{Coefficient} & \textbf{Prior} & \textbf{Coefficient} & \textbf{Prior}
\\ \hline
\multicolumn{2}{c}{$w_{DE}\left( z\right) $} & \multicolumn{2}{c}{$\gamma
\left( z\right) $} \\ \hline
$c_{0}^{w}$ & $\mathcal{N}(-1,\,0.5^{2})$ & $c_{0}^{\gamma }$ & $\mathcal{N}%
(0,\,1^{2})$ \\ 
$c_{1}^{w}$ & $\mathcal{N}(0,\,0.25^{2})$ & $c_{1}^{\gamma }$ & $\mathcal{N}%
(0,\,0.25^{2})$ \\ 
$c_{2}^{w}$ & $\mathcal{N}(0,\,0.167^{2})$ & $c_{2}^{\gamma }$ & $\mathcal{N}%
(0,\,0.167^{2})$ \\ 
$c_{3}^{w}$ & $\mathcal{N}(0,\,0.125^{2})$ & $c_{3}^{\gamma }$ & $\mathcal{N}%
(0,\,0.125^{2})$ \\ 
$c_{4}^{w}$ & $\mathcal{N}(0,\,0.1^{2})$ & $c_{N>3}^{\gamma }$ & $0$ \\ 
$c_{5}^{w}$ & $\mathcal{N}(0,\,0.083^{2})$ &  &  \\ 
$c_{N>5}^{w}$ & $0$ &  &  \\ \hline\hline
\end{tabular}%
\label{priors2}%
\end{table}%

\section{Learning $w_{DE}(z)$ and $\protect\gamma (z)$}

\label{sec5}

In this section, we present the results obtained from the training of the
Cosmo-PINN on the reconstruction of the dark energy equation of state
parameter $w_{DE}\left( z\right) $ and growth index $\gamma \left( z\right) $%
. The training is performed for the following combinations of data sets:
CC+BAO+RSD and~CC+BAO+SNIa+RSD, where SNIa corresponds to the PP, U3, DD and
PP+SH0ES catalogues.

\subsection{CC+BAO+RSD}

\begin{figure}[htbp]
\centering\includegraphics[width=1\textwidth]{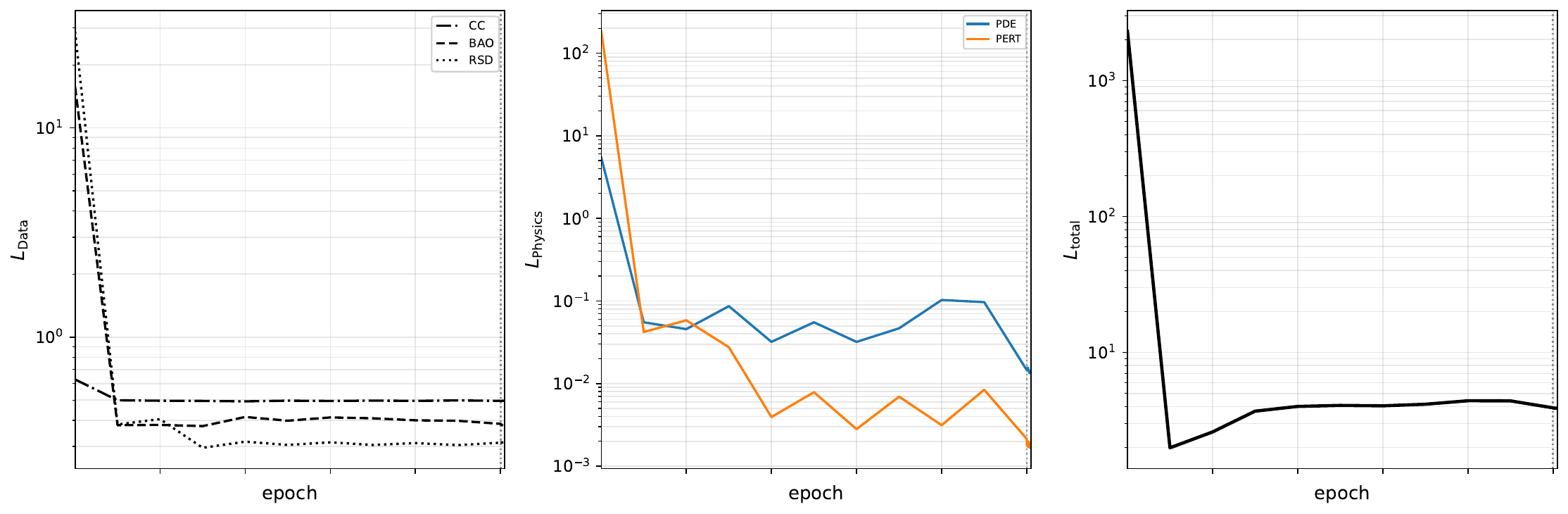}
\caption{No SNIa: Evolution of the loss components during training for the
combined data CC+BAO+RSD. The left panel corresponds to the data loss terms.
The middle panel shows the physics residuals, and the right panel the total
loss function. The vertical dotted line describes the transition to the
L-BFGS optimizer stage. }
\label{fig2}
\end{figure}

\begin{figure}[htbp]
\centering\includegraphics[width=0.8\textwidth]{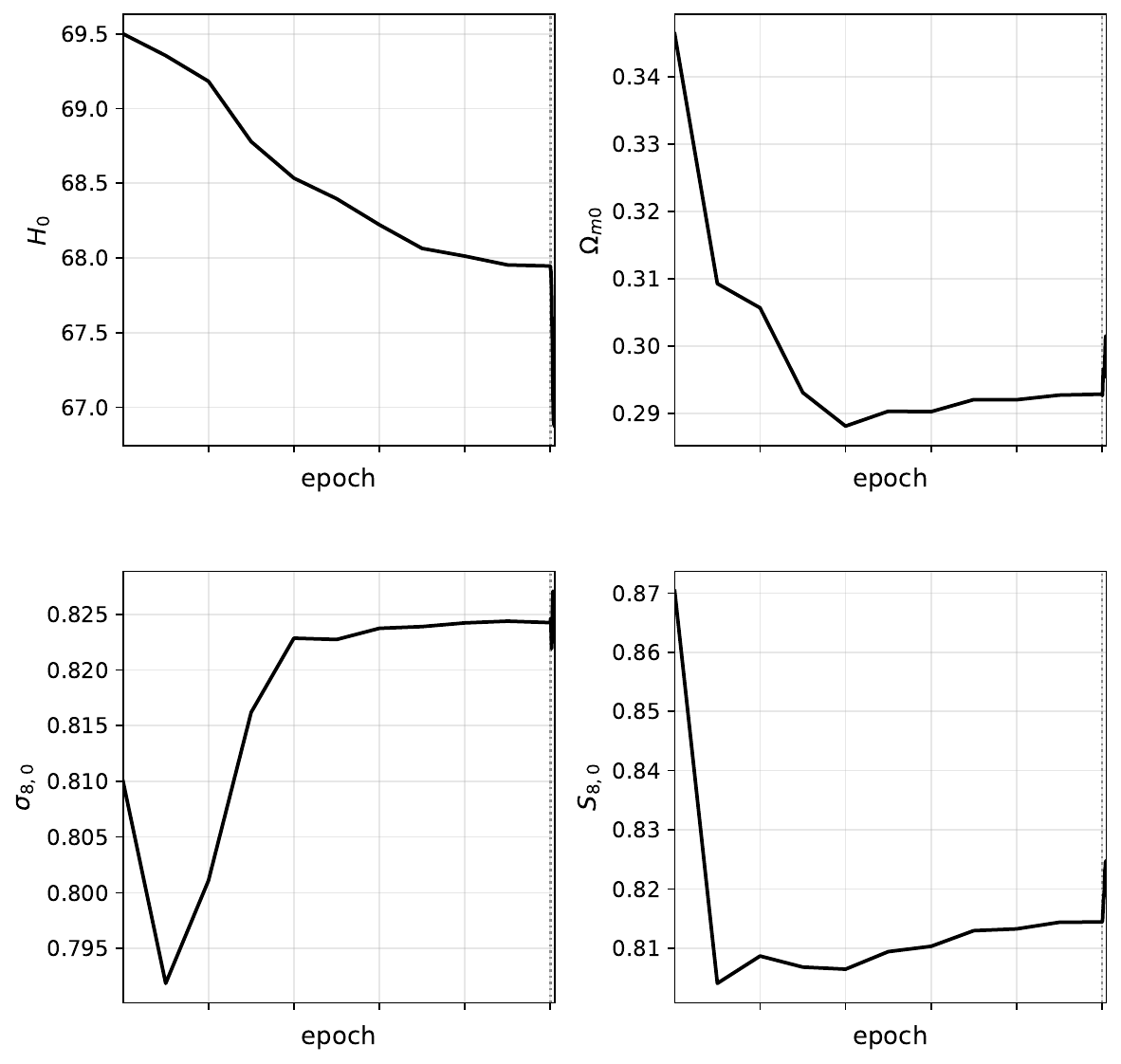}
\caption{No SNIa: Evolution of the cosmological parameters $H_{0},~\Omega
_{m0}$,~$\protect\sigma _{8,0}$ and of$~S_{8}$, during the Adam and L-BFGS
optimizer training for the combined data CC+BAO+RSD. The vertical dotted
line describes the transition to the L-BFGS optimizer stage. }
\label{fig4}
\end{figure}

\begin{figure}[tbph]
\centering\includegraphics[width=0.5\textwidth]{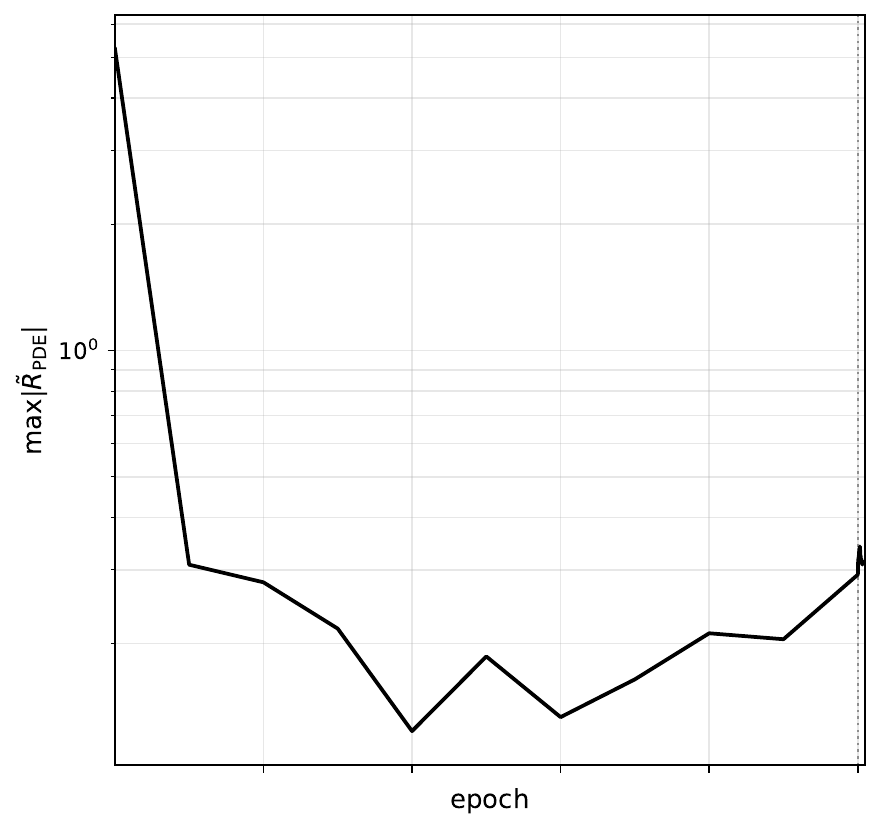}
\caption{No SNIa: Maximum absolute value of the PDE residual $\max
\left\vert \mathcal{R}_{PDE}\right\vert $ during training for the combined
data CC+BAO+RSD. The vertical dotted line describes the transition to the
L-BFGS optimizer stage. }
\label{fig3a}
\end{figure}

We train the Cosmo-PINN by using RSD data combined with the CC and BAO\
measurements. Fig. \ref{fig2} shows the evolution of the loss components
during the first two stages of the training, that is, for the Adam-optimizer
and L-BFGS optimizer epochs. We present the loss components for the
background data, $\mathcal{L}_{DATA}$, and the RSD measurements $\mathcal{L}%
_{RSD}$, the residuals $\mathcal{L}_{PDE},~\mathcal{L}_{PERT}$ which enforce
the physical laws within the PINN, and the total loss function $\mathcal{L}%
_{total}$. \ The loss components decrease rapidly during the Adam stage,
which means that the Cosmo-PINN fits the observational data with a solution
that satisfies the physics residuals. Finally, the L-BFGS stage refines and
further constrains the cosmological parameters $H_{0},~\Omega _{m0}$ and $%
\sigma _{8,0}$ as shown in Fig. \ref{fig4}. Furthermore, in Fig. \ref{fig3a}
we plot the maximum absolute value of the PDE residual $\max \left\vert 
\mathcal{R}_{PDE}\right\vert $ for the Adam-optimizer and L-BFGS optimizer
epochs, from which we conclude that the reconstructed solution satisfies the
physical law across the training period for the whole redshift range.

\begin{figure}[tbph]
\centering\includegraphics[width=0.5\textwidth]{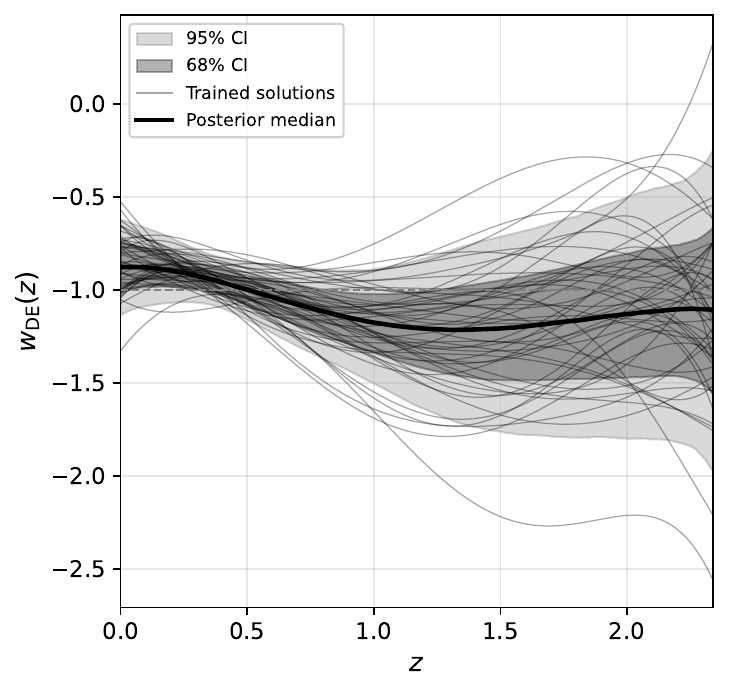}
\caption{No SNIa: Reconstructed dark energy equation of state parameter $%
w_{DE}\left( z\right) ~$with the $68\%$ and $95\%~$CI and individual trained
solutions. The reconstruction is for the combined data CC+BAO+RSD.}
\label{fig5}
\end{figure}

\begin{figure}[tbph]
\centering\includegraphics[width=0.7\textwidth]{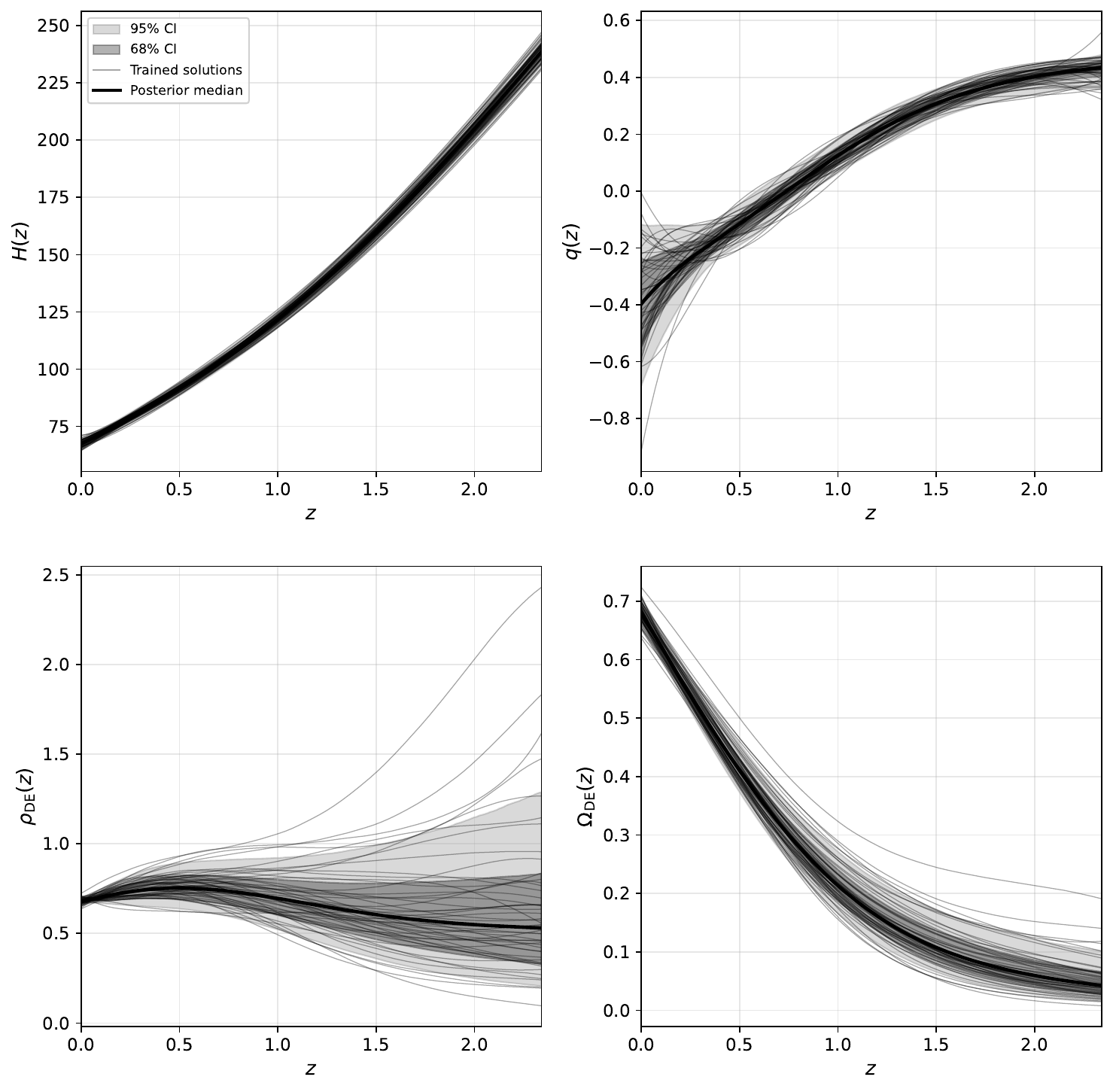}
\caption{No SNIa: Reconstructed Hubble function $H\left( z\right) $,
deceleration parameter $q\left( z\right) $, dark energy density $\protect%
\rho _{DE}\left( z\right) $ and $\Omega _{DE}\left( z\right) ~$with the $%
68\% $ and $95\%~$CI and individual trained solutions. The reconstruction is
for the combined data CC+BAO+RSD.}
\label{fig6}
\end{figure}

In Fig. \ref{fig5} we present the reconstructed dark energy equation of
state parameter $w_{DE}\left( z\right) $ with the $68\%$ and $95\%~$credible
intervals (CI) as given by the Bayesian stage. For illustration, we present
some individual trained solutions which were used to construct the
posterior.\ The posterior median suggests a present value $w_{DE}\left(
z\rightarrow 0\right) >-1$, while $w_{DE}~$crosses the phantom divide line
at $z\simeq 0.5 $. Moreover, Fig. \ref{fig6} summarizes the reconstructed
evolution of the Hubble function $H\left( z\right) $, the deceleration
parameter $q\left( z\right) =-1+\left( 1+z\right) \frac{d\ln H}{dz}$, and
the dark energy density parameter $\Omega _{DE}$. We also present the dark
energy density $\rho _{DE}\left( z\right) $, which is directly learned by
the network and is used to derive the rest of the cosmological quantities.

\begin{figure}[tbph]
\centering\includegraphics[width=0.9\textwidth]{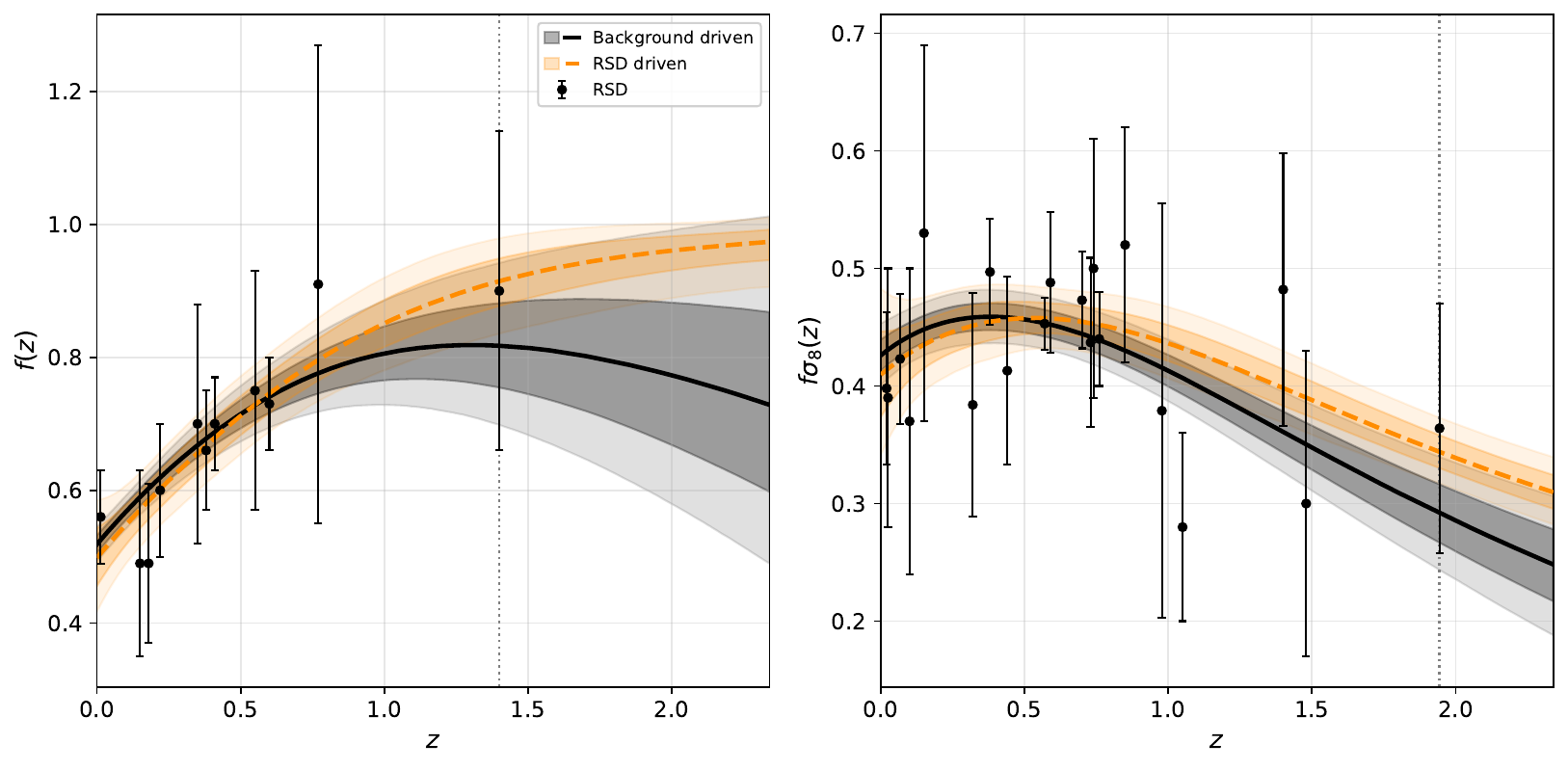}
\caption{No SNIa: Reconstructed $f\left( z\right) $ and $f\protect\sigma %
_{8}\left( z\right) $ from the $\protect\delta _{m}$ and $\protect\gamma (z)$
modules and the corresponding $68\%$ and $95\%$ CI. The reconstruction is
for the CC+BAO+RSD combined data. The vertical dotted line indicates the
maximum redshift at which RSD measurements are available.}
\label{fig7}
\end{figure}

As far as the growth rate $f(z)$ and $f\sigma _{8}(z)$ are concerned, in
Fig.~\ref{fig7} we show the results of the reconstruction from the two
modules, the growth $\delta _{m}$ module and the growth index $\gamma (z)$
module; the latter provides an individual reconstruction independent of the
perturbation equation. We observe that for $z<0.8$ the two bands overlap,
with the solution reconstructed by the $\delta _{m}$ module having tighter
bands. However, at higher redshifts there are fewer data points, which do
not allow for a proper reconstruction of the perturbation functions from the
RSD data alone. This is not the case for the solutions obtained from the
growth $\delta _{m}$ module, since the residual $\mathcal{R}_{PERT}(z)$
should always be satisfied, which drives the evolution of the perturbations.
In Tables \ref{table2} and \ref{table3} we present the observable values of $%
f\left( z\right) $ and $f\sigma _{8}\left( z\right) $ as given by the data
sets, and compare them with the reconstructed values through the two modules.

\begin{table}[tbp] \centering%
\caption{$f(z)$ measurements from the RSD data and the reconstructed
values from the growth $\delta_{m}$ and growth index $\gamma(z)$ modules. The reconstruction is for the combined data CC+BAO+RSD.}%
\begin{tabular}{cccc}
\hline\hline
$\mathbf{z}$ & $\mathbf{f}_{obs}\pm \mathbf{\sigma }$ & $\mathbf{f}$ via $%
\delta _{m}\left( z\right) $ & $\mathbf{f}$ via $\gamma \left( z\right) $ \\ 
\hline
$0.013$ & $0.56_{-0.07}^{+0.07}$ & $0.525_{-0.013}^{+0.013}$ & $%
0.505_{-0.040}^{+0.042}$ \\ 
$0.150$ & $0.49_{-0.14}^{+0.14}$ & $0.590_{-0.013}^{+0.013}$ & $%
0.571_{-0.028}^{+0.028}$ \\ 
$0.180$ & $0.49_{-0.12}^{+0.12}$ & $0.603_{-0.014}^{+0.013}$ & $%
0.584_{-0.028}^{+0.028}$ \\ 
$0.220$ & $0.60_{-0.10}^{+0.10}$ & $0.619_{-0.015}^{+0.014}$ & $%
0.602_{-0.028}^{+0.028}$ \\ 
$0.350$ & $0.70_{-0.18}^{+0.18}$ & $0.668_{-0.018}^{+0.017}$ & $%
0.656_{-0.029}^{+0.030}$ \\ 
$0.380$ & $0.66_{-0.09}^{+0.09}$ & $0.678_{-0.019}^{+0.018}$ & $%
0.668_{-0.029}^{+0.030}$ \\ 
$0.410$ & $0.70_{-0.07}^{+0.07}$ & $0.688_{-0.019}^{+0.019}$ & $%
0.679_{-0.029}^{+0.030}$ \\ 
$0.550$ & $0.75_{-0.18}^{+0.18}$ & $0.728_{-0.023}^{+0.023}$ & $%
0.729_{-0.029}^{+0.030}$ \\ 
$0.600$ & $0.73_{-0.07}^{+0.07}$ & $0.741_{-0.025}^{+0.025}$ & $%
0.746_{-0.029}^{+0.030}$ \\ 
$0.770$ & $0.91_{-0.36}^{+0.36}$ & $0.776_{-0.031}^{+0.031}$ & $%
0.796_{-0.031}^{+0.032}$ \\ 
$1.400$ & $0.90_{-0.24}^{+0.24}$ & $0.817_{-0.063}^{+0.064}$ & $%
0.915_{-0.038}^{+0.034}$ \\ \hline\hline
\end{tabular}%
\label{table2}%
\end{table}%

\begin{table}[tbp] \centering%
\caption{$f\sigma_8 (z)$ measurements from the RSD data and the
reconstructed values from the growth $\delta_{m}$ and growth index
$\gamma(z)$ modules. The reconstruction is for the combined data CC+BAO+RSD.}%
\begin{tabular}{cccc}
\hline\hline
$\mathbf{z}$ & $\mathbf{f\sigma }_{8,obs}\pm \sigma $ & $\mathbf{f\sigma }%
_{8}$ via $\delta _{m}\left( z\right) $ & $\mathbf{f\sigma }_{8}$ via $%
\gamma \left( z\right) $ \\ \hline
$0.020$ & $0.398_{-0.065}^{+0.065}$ & $0.430_{-0.013}^{+0.013}$ & $%
0.414_{-0.032}^{+0.033}$ \\ 
$0.025$ & $0.390_{-0.110}^{+0.110}$ & $0.431_{-0.013}^{+0.013}$ & $%
0.415_{-0.032}^{+0.032}$ \\ 
$0.067$ & $0.423_{-0.055}^{+0.055}$ & $0.438_{-0.012}^{+0.012}$ & $%
0.422_{-0.026}^{+0.026}$ \\ 
$0.100$ & $0.370_{-0.130}^{+0.130}$ & $0.442_{-0.012}^{+0.012}$ & $%
0.428_{-0.023}^{+0.023}$ \\ 
$0.150$ & $0.530_{-0.160}^{+0.160}$ & $0.448_{-0.011}^{+0.012}$ & $%
0.435_{-0.020}^{+0.020}$ \\ 
$0.320$ & $0.384_{-0.095}^{+0.095}$ & $0.458_{-0.011}^{+0.012}$ & $%
0.451_{-0.017}^{+0.017}$ \\ 
$0.380$ & $0.497_{-0.045}^{+0.045}$ & $0.459_{-0.011}^{+0.012}$ & $%
0.455_{-0.016}^{+0.016}$ \\ 
$0.440$ & $0.413_{-0.080}^{+0.080}$ & $0.458_{-0.011}^{+0.011}$ & $%
0.457_{-0.015}^{+0.015}$ \\ 
$0.570$ & $0.453_{-0.022}^{+0.022}$ & $0.453_{-0.011}^{+0.011}$ & $%
0.458_{-0.013}^{+0.013}$ \\ 
$0.590$ & $0.488_{-0.060}^{+0.060}$ & $0.452_{-0.011}^{+0.011}$ & $%
0.457_{-0.013}^{+0.013}$ \\ 
$0.700$ & $0.473_{-0.041}^{+0.041}$ & $0.444_{-0.012}^{+0.012}$ & $%
0.455_{-0.013}^{+0.013}$ \\ 
$0.730$ & $0.437_{-0.072}^{+0.072}$ & $0.442_{-0.012}^{+0.012}$ & $%
0.453_{-0.013}^{+0.013}$ \\ 
$0.740$ & $0.500_{-0.110}^{+0.110}$ & $0.441_{-0.012}^{+0.012}$ & $%
0.453_{-0.013}^{+0.013}$ \\ 
$0.760$ & $0.440_{-0.040}^{+0.040}$ & $0.439_{-0.012}^{+0.012}$ & $%
0.452_{-0.013}^{+0.014}$ \\ 
$0.850$ & $0.520_{-0.100}^{+0.100}$ & $0.430_{-0.012}^{+0.012}$ & $%
0.447_{-0.014}^{+0.014}$ \\ 
$0.978$ & $0.379_{-0.176}^{+0.176}$ & $0.416_{-0.013}^{+0.013}$ & $%
0.438_{-0.015}^{+0.015}$ \\ 
$1.050$ & $0.280_{-0.080}^{+0.080}$ & $0.407_{-0.014}^{+0.014}$ & $%
0.432_{-0.015}^{+0.016}$ \\ 
$1.400$ & $0.482_{-0.116}^{+0.116}$ & $0.361_{-0.018}^{+0.017}$ & $%
0.398_{-0.015}^{+0.016}$ \\ 
$1.480$ & $0.300_{-0.130}^{+0.130}$ & $0.350_{-0.019}^{+0.018}$ & $%
0.390_{-0.015}^{+0.016}$ \\ 
$1.944$ & $0.364_{-0.106}^{+0.106}$ & $0.292_{-0.025}^{+0.024}$ & $%
0.344_{-0.014}^{+0.014}$ \\ \hline\hline
\end{tabular}%
\label{table3}%
\end{table}%

\begin{figure}[tbph]
\centering\includegraphics[width=0.5\textwidth]{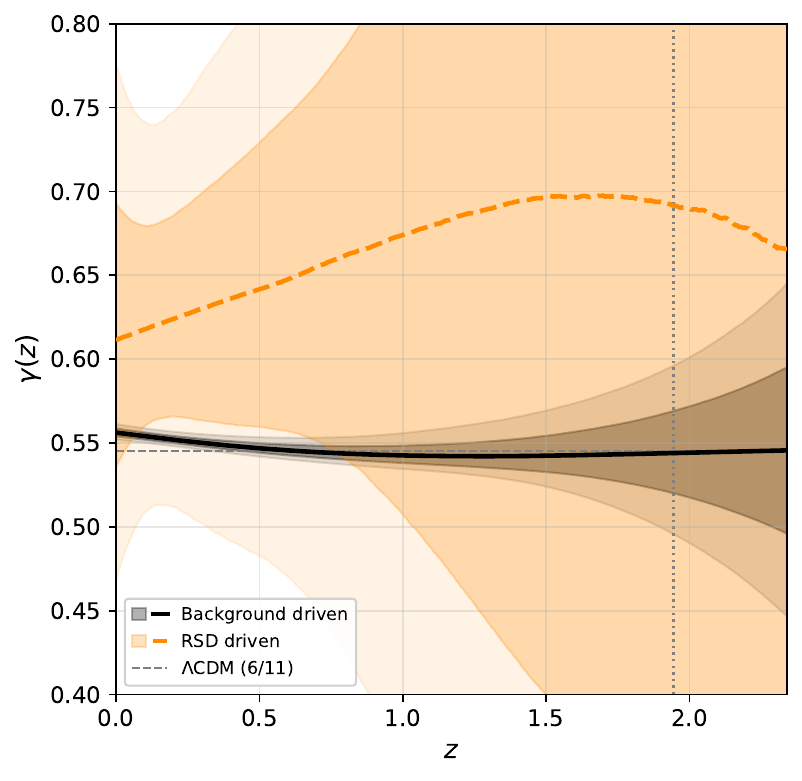}
\caption{No SNIa: We compare the individual reconstruction $\protect\gamma %
(z)$ with $\protect\gamma _{\mathrm{GR}}(z)$, derived from the $\protect%
\delta _{m}$ module, for the same reconstructed background. The
reconstruction is for the CC+BAO+RSD combined data. The vertical dotted line
indicates the maximum redshift at which RSD measurements are available.}
\label{fig8}
\end{figure}

The growth index $\gamma (z)$ obtained from the two different reconstruction
approaches is shown in Fig.~\ref{fig8}. We observe that the individual
reconstruction of $\gamma (z)$ has very wide posterior bands in comparison
to $\gamma _{GR}\left( z\right) $ obtained from the $\delta _{m}$~module.
This is because the growth index is not directly constrained by the data.
The growth rate $f(z)$ is constrained, which is related to $\gamma (z)$
through expression~(\ref{sc.05a}). Thus, at the matter epoch, $\Omega
_{m}\rightarrow 1$, the growth rate becomes independent of $\gamma (z)$ and
is driven to $f\rightarrow 1$, as can easily be seen from the reconstruction
in Fig.~\ref{fig7}. This limits the applicability of the growth index
formalism at high redshifts. For small redshifts, $z<0.8$, we observe that
the $\gamma _{GR}\left( z\right) $ lies outside the $68\%$~CI of the $\gamma
\left( z\right) $; nevertheless, it is always within the $95\%$ interval,
suggesting no tension between them. 

\begin{figure}[tbph]
\centering\includegraphics[width=0.8\textwidth]{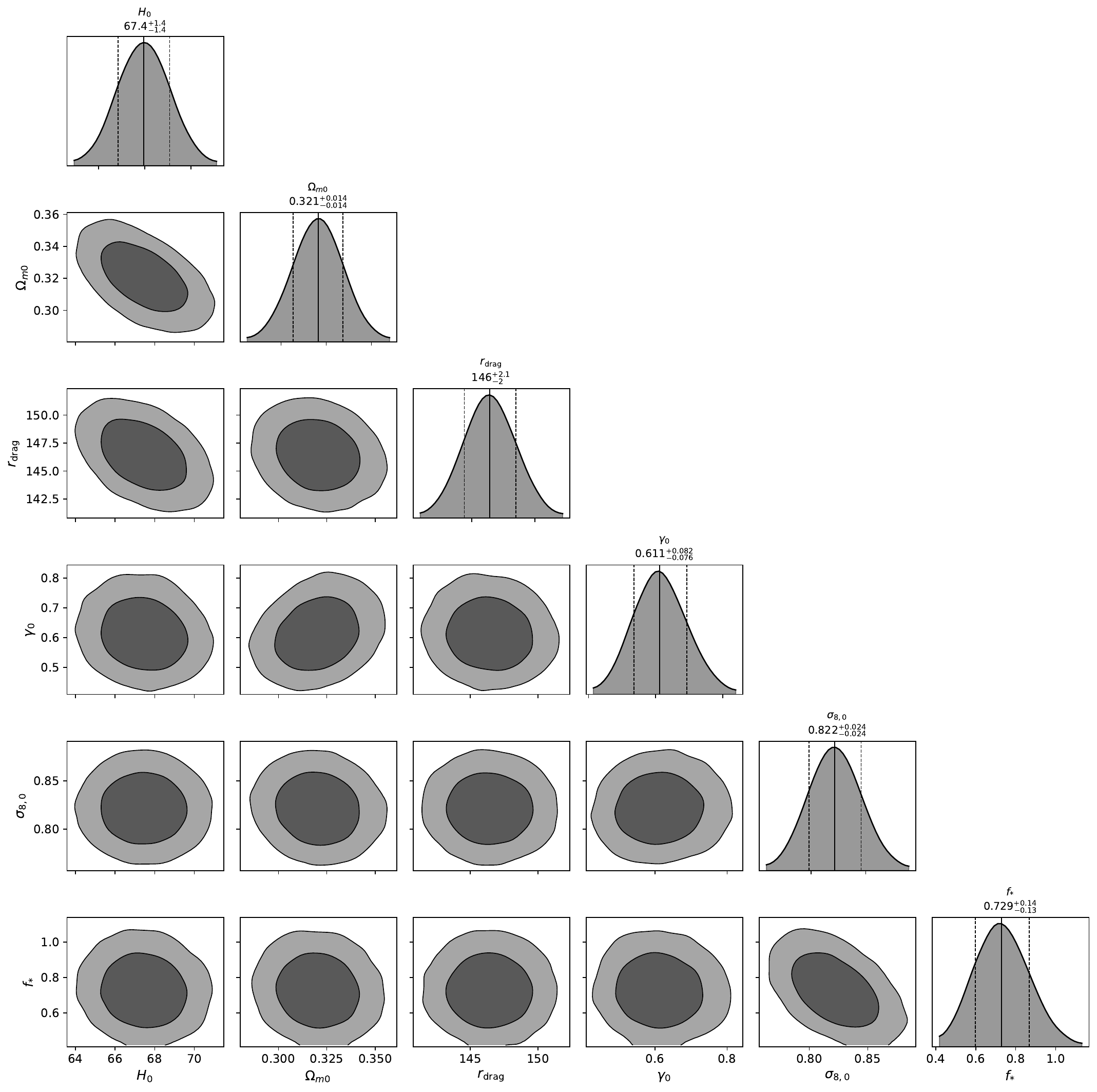}
\caption{No SNIa: Marginalized distributions for the cosmological parameters 
$H_{0}$, $\Omega _{m0}$, $\protect\gamma _{0}$, $\protect\sigma _{8,0}$ and $%
f_{\star }$, obtained from the HMC/NUTS analysis for the CC+BAO+RSD data.
The contours correspond to the $68\%$ and $95\%$ credible regions.}
\label{fig9}
\end{figure}

\begin{figure}[tbph]
\centering\includegraphics[width=0.8\textwidth]{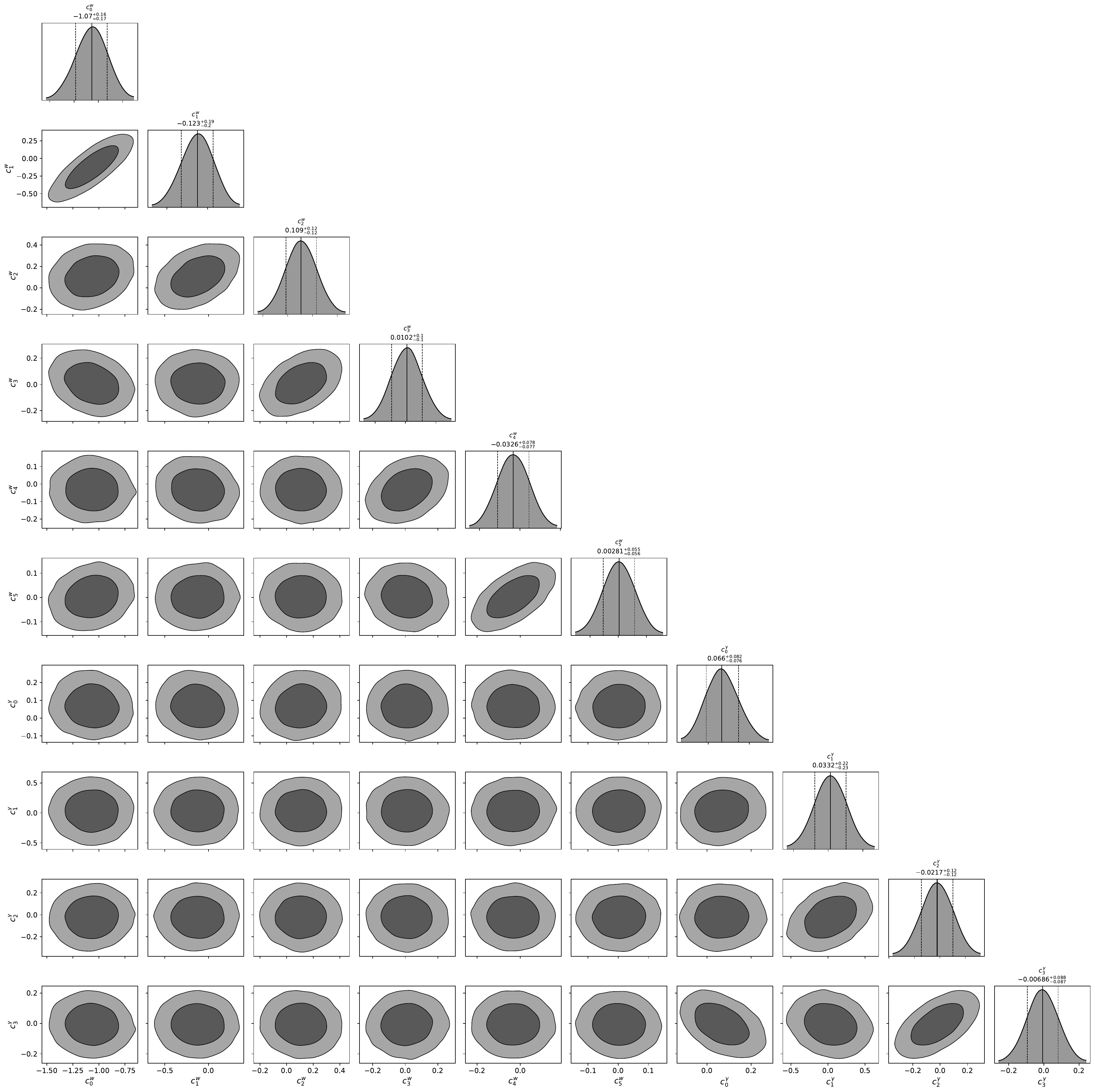}
\caption{No SNIa: Marginalized posterior distributions for the Chebyshev
coefficients $c_{N}^{w}$ and $c_{N}^{\protect\gamma }$ entering the
reconstruction of the dark energy equation of state parameter $w_{DE}\left(
z\right) $ and of the growth index $\protect\gamma \left( z\right) $,
obtained from the HMC/NUTS analysis for the CC+BAO+RSD data. The contours
correspond to the $68\%$ and $95\%$ credible regions.}
\label{fig10}
\end{figure}

Finally, in Figs.~\ref{fig9} and~\ref{fig10} we show the posterior
distributions of the trained cosmological parameters $H_{0}$, $\Omega _{m0}$%
, $\gamma _{0}$, $\sigma _{8,0}$, $f_{\star }$, and of the Chebyshev
coefficients $c_{N}^{w}$ and $c_{N}^{\gamma }$, respectively. We see that
the coefficients have weak degeneracies, which follows from the selection of
the shrinkage priors.

\subsection{CC+BAO+SNIa+RSD}

We extend the analysis by introducing the SNIa data in the training of the
Cosmo-PINN. For the combined data set CC+BAO+SNIa+RSD, Table~\ref{tabl4}
summarizes the posterior medians of the cosmological parameters $H_{0}$, $%
\Omega _{m0}$, $\sigma _{8,0}$, and $f_{\ast }$, and of the derived
parameters $S_{8}$ and $q\left( z\rightarrow 0\right) =q_{0}$. It also shows
the present-day values of the reconstructed equation of state $%
w_{DE}(z\rightarrow 0)=w_{DE,0}$, and of the growth index, $\gamma \left(
z\rightarrow 0\right) =\gamma _{0}$, as obtained from both the $\delta
_{m}(z)$ and $\gamma (z)$ modules.

The supernova data do not shift the median values of the cosmological
parameters significantly, but mainly tighten the CI. The Hubble constant has
median values within the range $H_{0}\in \left[ 67.24,69.65\right] $~$\frac{%
km}{sMpc}$, with the largest value obtained when the SH0ES calibration is
included. Furthermore, the energy density of the total matter today is
derived within the range $\Omega _{m0}\in $ $\left[ 0.3153,0.3232\right] $.
The parameters $\mathbf{\sigma }_{8,0}$ and $S_{8}$, related to the
amplitude of matter fluctuations and clustering, have posterior medians $%
\sigma _{8,0}\in \left[ 0.8185,0.8227\right] $ and $S_{8}\in \left[
0.8397,0.8534\right] $. The trained parameter $f_{\ast }$ is always found to
be lower than~one, and the median values are within the range $f_{\ast }\in %
\left[ 0.716,0.744\right] $.

As far as the reconstructed functions $w_{DE}(z)$ and $\gamma (z)$ are
concerned, the posterior medians of $w_{DE,0}~$lie in the quintessence
regime, $w_{DE,0}\in \left[ -0.930,-0.867\right] $, with the lowest value
obtained when the SH0ES calibration is introduced. The growth index is
tightly constrained when it is derived via the $\delta _{m}(z)$ module, with 
$\gamma _{GR,0}\simeq 0.555$ and very small uncertainties for all
combinations. This is because, in this approach, $\gamma _{GR}(z)$ follows
from the solution of the perturbation equation of General Relativity and
thus it is fully determined by the reconstructed background. On the other
hand, when $\gamma _{0}$ is determined directly from the RSD data via the $%
\gamma (z)$ module, the obtained values lie in the range $\gamma _{0}\in %
\left[ 0.601,0.616\right] $, with the $\Lambda $CDM limit $\frac{6}{11}$
always within the $68\%$ CI. The obtained values indicate a suppression of
the growth of cosmic structure, in agreement with the analysis presented in 
\cite{prl1}. However, its statistical significance is low, which means that
the RSD data do not support a departure from General Relativity.

\begin{figure}[tbph]
\centering\includegraphics[width=0.8\textwidth]{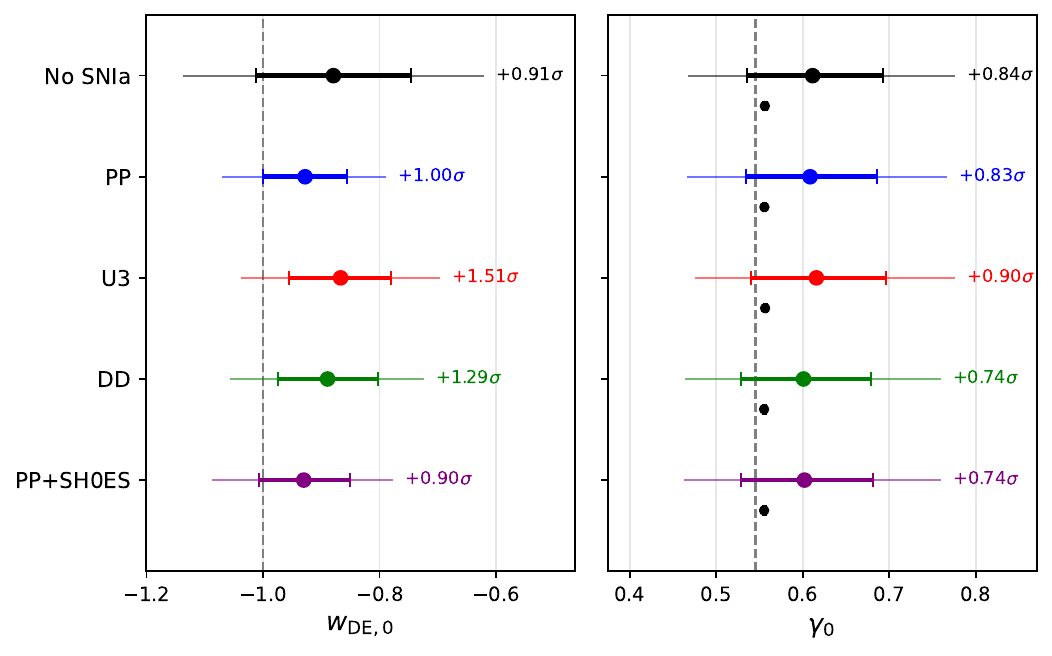}
\caption{Comparison of the posterior median and the $68\%$ CI for the
cosmological parameters as provided by the training of the Cosmo-PINN and
summarized in Table \protect\ref{tabl4}. The left panel shows $w_{DE,0}$,
and the right panel the present value of $\protect\gamma _{0}$. The dots
show the posterior median of $\protect\gamma _{0}$ as provided by General
Relativity for the same background.}
\label{fig11}
\end{figure}

The results are visualized in Fig. \ref{fig11}, from which we remark the
consistency of the reconstruction of the cosmological parameters,
independently of the supernova catalogue that is applied.

\begin{table}[tbp] \centering%
\caption{Posterior median values and $68\%$ credible intervals for the
cosmological parameters obtained from the HMC analysis for five different
dataset combinations.} 
\begin{tabular}{cccccc}
\hline\hline
\textbf{Dataset} & \textbf{no SNIa} & \textbf{PP} & \textbf{U3} & \textbf{DD}
& \textbf{PP+SH0ES} \\ \hline
$\mathbf{H}_{0}$ & $67.45_{-1.39}^{+1.40}$ & $67.74_{-1.00}^{+1.00}$ & $%
67.24_{-1.08}^{+1.08}$ & $67.88_{-1.00}^{+1.03}$ & $69.65_{-0.84}^{+0.84}$
\\ 
$\mathbf{\Omega }_{m0}$ & $0.3205_{-0.0139}^{+0.0138}$ & $%
0.3178_{-0.0107}^{+0.0099}$ & $0.3232_{-0.0116}^{+0.0109}$ & $%
0.3165_{-0.0106}^{+0.0101}$ & $0.3153_{-0.0104}^{+0.0095}$ \\ 
$\mathbf{q}_{0}$ & $-0.397_{-0.147}^{+0.145}$ & $-0.451_{-0.076}^{+0.074}$ & 
$-0.380_{-0.094}^{+0.092}$ & $-0.412_{-0.090}^{+0.089}$ & $%
-0.456_{-0.081}^{+0.082}$ \\ 
$\mathbf{\sigma }_{8,0}$ & $0.8216_{-0.0237}^{+0.0242}$ & $%
0.8198_{-0.0231}^{+0.0236}$ & $0.8227_{-0.0236}^{+0.0241}$ & $%
0.8185_{-0.0231}^{+0.0247}$ & $0.8196_{-0.0230}^{+0.0241}$ \\ 
$\mathbf{S}_{8,0}$ & $0.8488_{-0.0289}^{+0.0300}$ & $%
0.8433_{-0.0258}^{+0.0260}$ & $0.8534_{-0.0270}^{+0.0275}$ & $%
0.8399_{-0.0254}^{+0.0270}$ & $0.8397_{-0.0255}^{+0.0265}$ \\ 
$\mathbf{f}_{\ast }$ & $0.729_{-0.131}^{+0.140}$ & $0.744_{-0.128}^{+0.139}$
& $0.716_{-0.129}^{+0.138}$ & $0.738_{-0.127}^{+0.142}$ & $%
0.738_{-0.128}^{+0.142}$ \\ 
$\mathbf{w}_{DE,0}$ & $-0.879_{-0.133}^{+0.133}$ & $-0.928_{-0.073}^{+0.071}$
& $-0.867_{-0.089}^{+0.087}$ & $-0.889_{-0.086}^{+0.086}$ & $%
-0.930_{-0.077}^{+0.079}$ \\ 
$\gamma _{GR,0}$ & $0.5562_{-0.0022}^{+0.0024}$ & $%
0.5556_{-0.0014}^{+0.0015} $ & $0.5565_{-0.0016}^{+0.0018}$ & $%
0.5553_{-0.0013}^{+0.0016}$ & $0.5553_{-0.0013}^{+0.0016}$ \\ 
$\gamma _{0}$ & $0.611_{-0.076}^{+0.082}$ & $0.609_{-0.074}^{+0.078}$ & $%
0.616_{-0.076}^{+0.080}$ & $0.601_{-0.072}^{+0.078}$ & $%
0.602_{-0.073}^{+0.079}$ \\ \hline\hline
\end{tabular}%
\label{tabl4}%
\end{table}%

\begin{figure}[tbph]
\centering\includegraphics[width=0.8\textwidth]{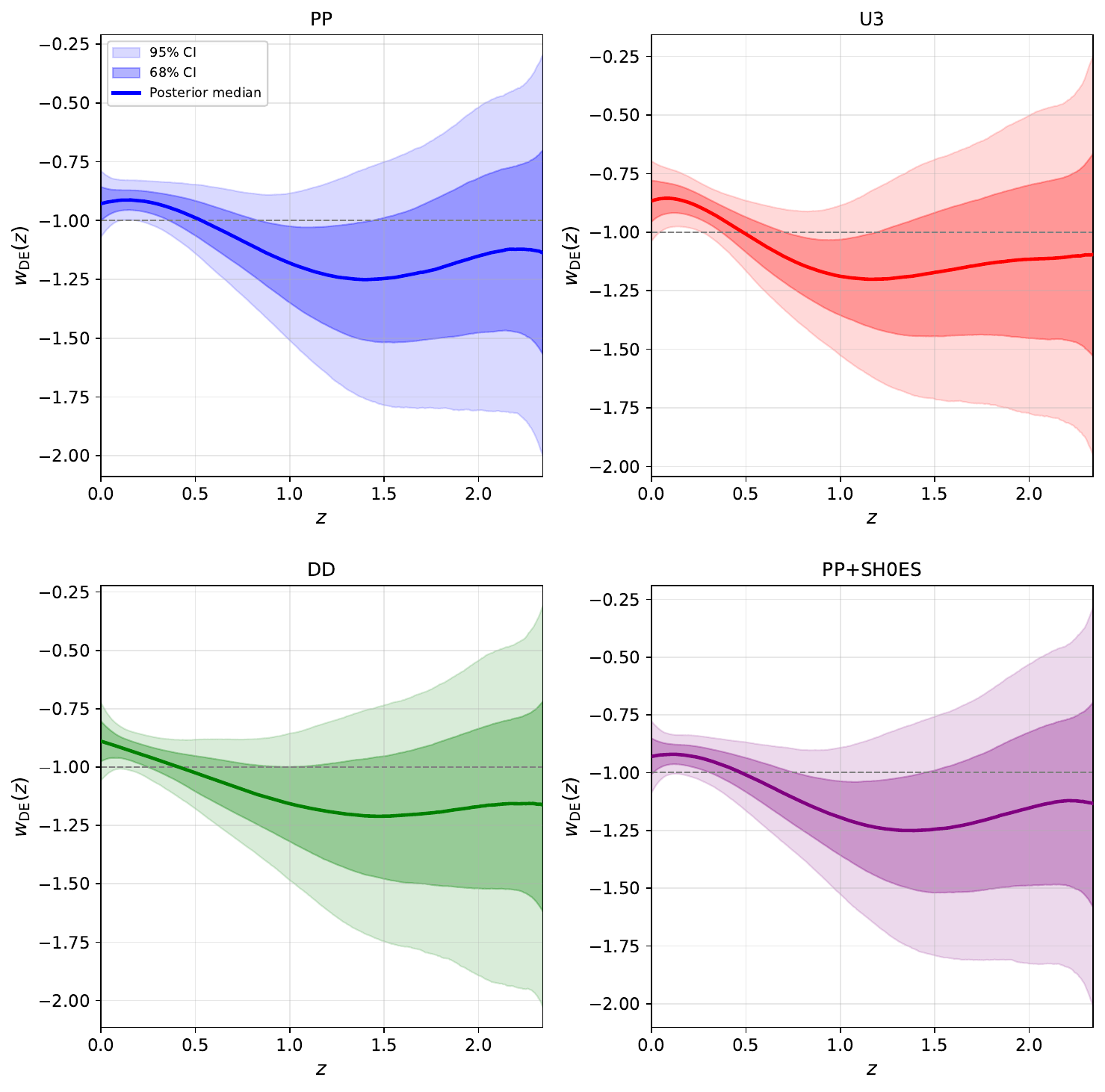}
\caption{CC+BAO+SNIa+RSD: Reconstructed dark energy equation of state
parameter $w_{DE}\left( z\right) ~$with the $68\%$ and $95\%~$CI and
individual trained solutions. The reconstruction is for the combined data
CC+BAO+SNIa+RSD.}
\label{fig12}
\end{figure}

\begin{figure}[tbph]
\centering\includegraphics[width=0.8\textwidth]{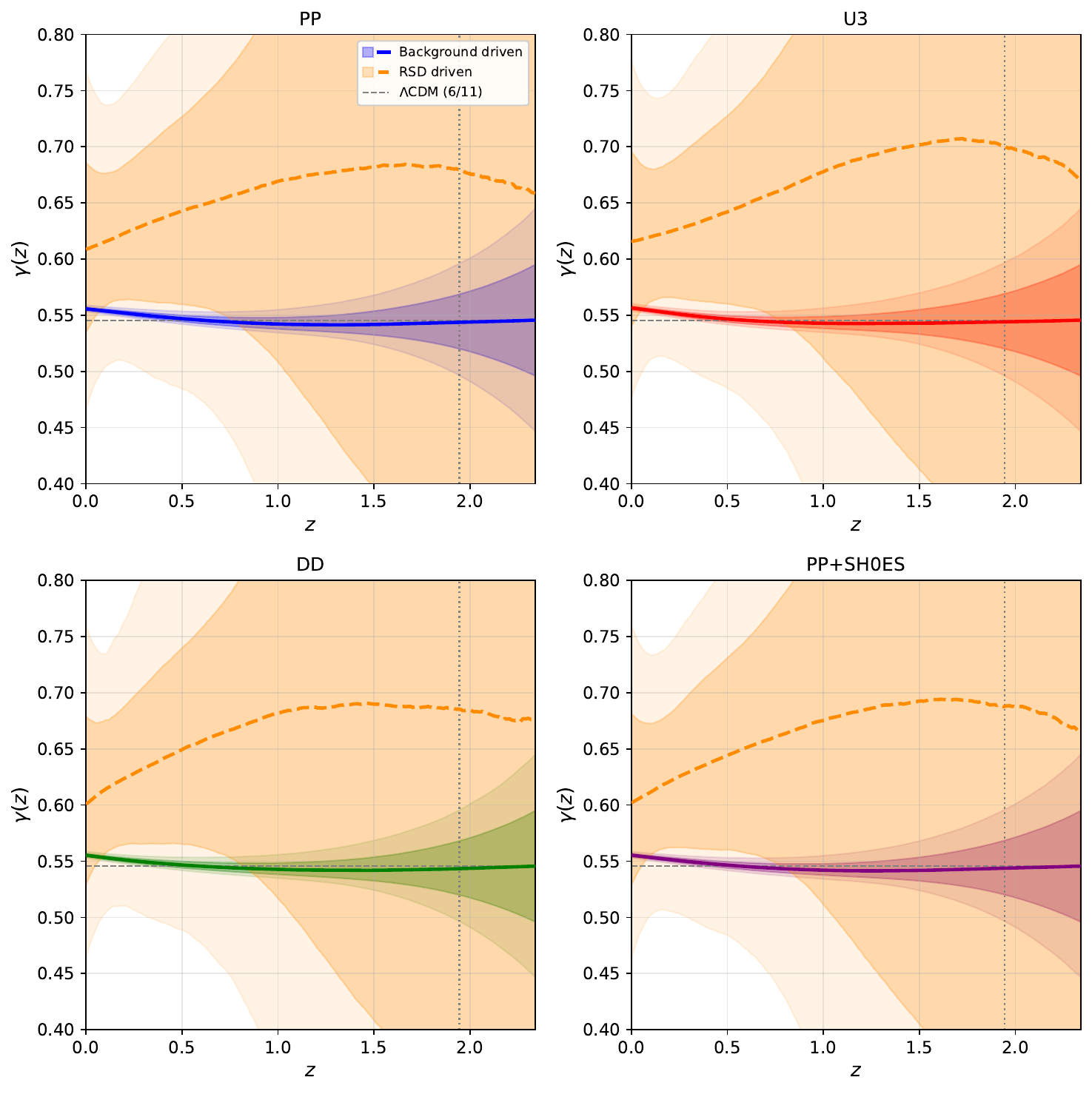}
\caption{CC+BAO+SNIa+RSD: Individual reconstruction of $\protect\gamma (z)$
and $\protect\gamma _{\mathrm{GR}}(z)$ for the same reconstructed
background. The reconstruction is for the CC+BAO+SNIa+RSD combined data. The
vertical dotted line indicates the maximum redshift at which RSD
measurements are available.}
\label{fig14}
\end{figure}

The reconstructed background function $w_{DE}\left( z\right) $ for the four
combinations is presented in Fig. \ref{fig12}, while in Fig. \ref{fig14} we
show the reconstructed $\gamma \left( z\right) $ functions obtained by the
two modules with their credible intervals.

For the different catalogues of SNIa, the $w_{DE}\left( z\right) $ today is
within the quintessence regime, and it crosses the phantom divide line for
all the datasets. Without SNIa data, the median $w_{DE}\left( z\right) $
takes the value $-1$ at $z=0.503$; for the PP catalogue, the crossing occurs
at $z=0.533$, while the SH0ES calibration shifts the crossing to $z=0.475$.
Moreover, for the U3 catalogue the $w_{DE}\left( z\right) $ crosses the
phantom divide line at $z=0.483$, while the DD catalogue supports a later
crossing at $z=0.412$. Finally, for the growth index, the SNIa modify the
dark energy equation of state parameter which affects the reconstructed $%
\gamma \left( z\right) $; however, the conclusions drawn before without SNIa
data remain unchanged.

\section{Conclusions}

\label{sec6}

The Cosmo-PINN framework has been extended through the incorporation of
modules that incorporate the RSD measurements into the data-driven
reconstruction, allowing for the reconstruction of the growth index of
matter perturbations. Specifically, the reconstructed background quantities,
derived from $\rho _{DE}(z)$ and $w_{DE}(z)$, are shared with two modules.
In the matter growth module, the network learns the growth of matter $\delta
_{m}(z)$, with the second-order linear perturbation equation of General
Relativity introduced as a residual term in the loss function. This module
provides the observables $f(z)$ and $f\sigma _{8}(z)$, which are constrained
by the RSD data, while the General Relativity prediction for the growth
index, $\gamma _{GR}(z)$, is also calculated. On the other hand, in the
growth index module, the network learns $\gamma (z)$ directly from the RSD
data, by using the already reconstructed background. This double
reconstruction of the growth index allows us to perform a consistency test
of General Relativity. The posterior distributions of the reconstructed
functions and of the cosmological parameters were derived through HMC
sampling around the MAP solution.

We trained the neural network by combining the RSD measurements with
background data, specifically the CC, the DESI DR2 BAO, and three SNIa
compilations, with the PP sample considered both with and without the SH0ES
calibration. For all data set combinations employed in this work, the
physics residuals reach small values across the entire redshift domain of
the reconstruction, from which we conclude that the reconstructed solutions
satisfy the differential equations that describe the physical laws. The
outcome of the training is the dark energy density $\rho _{DE}(z)$, which
defines the Hubble function, the dark energy equation of state parameter $w_{%
\mathrm{DE}}(z)$, the growth of matter $\delta _{m}(z)$, and the growth
indices $\gamma _{GR}(z)$ and $\gamma (z)$ provided by the two modules.
Furthermore, the posterior distributions of the cosmological parameters $%
H_{0}$, $\Omega _{m0}$, $\sigma _{8,0}$, and $r_{\mathrm{drag}}$ are derived
from the HMC sampling. Although a Chebyshev expansion was considered for the
representation of $w_{DE}(z)$ and $\gamma (z)$, the reconstruction should be
seen as non-parametric, since the main network learns $\rho _{DE}(z)$, and
the Chebyshev expansion represents the solution for $w_{DE}(z)$ that follows
from the residual. Moreover, as demonstrated in~\cite{pinnan}, replacing the
Chebyshev expansion of $w_{DE}(z)$ with a second neural network yields
statistically equivalent results, with a slight increase in the
uncertainties but a reduced computational cost.

From the analysis of the different data set combinations, we conclude that
the introduction of the SNIa data slightly changes the median values of the
background parameters, while it tightens their posteriors. In particular, we
found $H_{0}\simeq \left[ 67.24,69.65\right] $~$\frac{km}{sMpc}$, which
increases to $H_{0}=69.65_{-0.84}^{+0.84}\frac{km}{sMpc}$, when the SH0ES
calibration is included. For the matter density, we find $\Omega _{m0}\simeq
0.32$, and $\sigma _{8,0}\simeq 0.82$, such that the clustering parameter is 
$S_{8,0}\simeq 0.84$, independently of the supernova catalogue. Moreover,
the value of the dark energy equation of state parameter at present is found
to be within the quintessence regime, $w_{DE,0}\in \lbrack -0.930,-0.867]$,
while the posterior median of $w_{DE}(z)$ crosses the phantom divide line at 
$z\simeq 0.41-0.53$. As far as the two independent reconstructions of the
growth index are concerned, the growth index derived from the matter growth
module is tightly constrained, $\gamma _{GR,0}\simeq 0.555$, since it is
fully determined by the reconstructed background. In contrast, the growth
index reconstructed directly from the RSD data has median values $\gamma
_{0}\simeq 0.60$--$0.62$, with significantly larger uncertainties. The two
reconstructions are consistent within the $68\%$ CI at the present epoch,
while for $z<0.8$, $\gamma _{GR}(z)$ lies outside the $68\%$ but always
within the $95\%$ credible interval of $\gamma (z)$. The preference of the
RSD data for $\gamma _{0}>6/11$ is in agreement with results obtained
previously \cite{prl1}. Furthermore, the growth index is weakly constrained
at high redshifts, since $f\rightarrow 1$ as $\Omega _{m}\rightarrow 1$
independently of $\gamma (z)$, which limits the applicability of the
growth-index formalism during the matter-dominated era. Future measurements
will significantly improve the constraints on $\gamma (z)$, allowing for a
more stringent test of General Relativity.

In future work, we plan to extend the Cosmo-PINN to modified theories of
gravity and to interacting dark sector models.

\textbf{Data Availability Statement:} The Cosmo-PINN code is available upon
reasonable request.

\begin{acknowledgments}
The author acknowledges support from FONDECYT Grant 1240514. AP acknowledges
the COST Action CA23130 \textquotedblleft Bridging high and low energies in
search of quantum gravity (BridgeQG)\textquotedblright .
\end{acknowledgments}

\end{document}